\documentclass[10pt,a4paper]{article}
\usepackage{jheppub}
\usepackage[utf8]{inputenc}
\usepackage[english]{babel}
\usepackage{bbm} 
\usepackage{amsmath}
\usepackage{amsfonts}
\usepackage{amssymb}
\usepackage{comment}

\usepackage{environ}
\usepackage{amsthm}

\newcommand{\mE}[1]{\ensuremath{\mathbb E_{#1}}}

\newcommand{\sgn}{\text{sign\,}}
\newcommand{\erf}{\text{erf}}

\NewEnviron{es}{
	\begin{equation}\begin{split}
			\ensuremath{\BODY}
	\end{split}\end{equation}
}

\newcommand{\bb}[1]{\boldsymbol{#1}}

\newtheorem*{Wrk}{Working assumption}
\newtheorem{Thm}{Theorem}
\newtheorem{Cor}{Corollary}
\newtheorem{Def}{Definition}
\newtheorem{Rmk}{Remark}
\newtheorem{Prp}{Proposition}
\newtheorem{Lem}{Lemma}

\title{A spectral approach to Hebbian-like neural networks}
\author[a,b]{Elena Agliari,}
\author[a,b]{Alberto Fachechi,}
\author[a]{Domenico Luongo}
\affiliation[a]{Dipartimento di Matematica “Guido Castelnuovo”, Sapienza Università di Roma, Roma, Italy}
\affiliation[b]{GNFM-INdAM, Gruppo Nazionale di Fisica Matematica, Istituto Nazionale di Alta Matematica,
	Italy}
\emailAdd{agliari@mat.uniroma1.it}
\abstract{We consider the Hopfield neural network as a model of associative memory and we define its neuronal interaction matrix $\bb J$ as a function of a set of $K \times M$ binary vectors $\{\bb \xi^{\mu, A} \}_{\mu=1,...,K}^{A=1,...,M}$ representing a sample of the reality that we want to retrieve. In particular, any item $\bb \xi^{\mu, A}$ is meant as a corrupted version of an unknown ground pattern $\bb \zeta^{\mu}$, that is the target of our retrieval process. We consider and compare two definitions for $\bb J$, referred to as supervised and unsupervised, according to whether the class $\mu$, each example belongs to, is unveiled or not, also, these definitions recover the paradigmatic Hebb's rule under suitable limits. The spectral properties of the resulting matrices are studied and used to inspect the retrieval capabilities of the related models as a function of their control parameters. 
}
\begin{document}
\maketitle

\section{Introduction}
Since its introduction, in the eighties, the Hopfield neural network has attracted a big deal of attention from a broad community of scientists at the interface of physics, mathematics and computer science \cite{hopfield1982neural,hopfield1985neural}.  In fact, the Hopfield network is recognized as a paradigmatic model for associative memory: if properly designed, it can store and retrieve a set of $K$ information patterns $\bb \xi = \{\bb \xi^\mu\}_{\mu=1}^K$, with $\bb \xi^\mu \in \{-1, +1\}^N$.
More precisely, the model consists of a set of $N$ binary neuron, whose configuration is denoted as $\bb \sigma \in \{-1, +1\}^N$, interacting pairwise and symmetrically by an interaction strength encoded by the coupling matrix $\bb J \in \mathbb R^{N\times N}$, and evolving in time in such a way that any neuron $\sigma_i$ gets progressively aligned with the local field $(\bb J \bb \sigma^T)_i$ acting on it and stemming from the neighbouring neurons. 
 The key point for the functioning of the model as an associative memory is therefore to design $\bb J$ in such a way that stored patterns are associated to attractors in the configuration space. The standard choice is inspired by Hebb's principle \cite{Hebb-1949} and reads as $\bb J = \bb \xi^T \bb \xi$, which, in the case of Rademacher patterns and in the large-size limit, ensures a storage capacity of approximately $0.14 N$ patterns, see e.g. \cite{Amit}.

 In this context, determining the structure of the attraction basins is a paramount goal in order to understand the information processing principles lying behind the associative memory functionalities and possibly to highlight qualitatively-different working regimes of the model corresponding to different parameter settings.
 Such analysis can be naturally framed by means of the statistical mechanics of spin glasses, as pioneered by Amit, Gutfreund and Sompolinsky \cite{AGS1,AGS2}. 
 On the mathematical side, several results have been derived for the Hopfield model, by exploiting different techniques -- ranging from large deviation analysis \cite{bovier1995large,bovier1996almost} to Guerra's interpolation 
 \cite{barra2012glassy,agliari2020generalized,AABO-JPA2020} -- and leading to bounds on the storage capacity \cite{bovier1999sharp,feng2001critical,loukianova1997lower,newman1988memory,lowe1998storage,bovier1992rigorous,bovier1993rigorous,baldi1987number,bovier2001spin}.
 %
The analytical investigations underlying these results have significantly benefited from the simple expression of the Hebbian interaction matrix. On the other hand, the cost for this simplicity is a limited capacity of the network: in the limit of large size $N$, a symmetric neural network can store up to $N$ patterns \cite{Gardner-JPA1988}, that is much higher than the aforementioned $0.14 N$. In fact, when the number of stored patterns gets larger, the related attraction basins tend to overlap, giving rise to frustration and, consequently, to a plethora of spurious attractors, whose retrieval is interpreted as an error of the network. Thus, several algorithms have been developed to optimize the model coupling matrix, enhancing the attractive power of stored patterns and increasing the critical storage capacity \cite{hopfield1983unlearning,personnaz1985information,kanter1987associative,plakhov1992modified,plakhov1994converging,plakhov1995convergent,van1997hebbian,horas1998unlearning,dotsenko1991statistical,dotsenko1991replica}; 
in general, the core idea of these algorithms is to modify the structure of the matrix $\bb J$, in order to disentangle the attraction basins of the patterns and then downsizing the harmful effect associated to the presence of spurious attractors. 
Among these algorithms, we recall the so-called \emph{dreaming} kernel \cite{fachechi2019dreaming,agliari2019dreaming,fachechi2022outperforming}, which shall constitute the starting point for our work.
\par\medskip
Beyond these variations on Hebb's theme, more recently, much attention has been devoted to a scenario in which the information supplied and used to build the interaction matrix does not correspond to \emph{ground-truth} patterns, but rather to \emph{examples} of theirs, namely to examples of the reality that we want to retrieve and hereafter referred to as $\{\bb \xi^{\mu,A}\}_{\mu=1,...,P}^{A=1,...,M}$ \cite{Fontanari-1990,AABD-NN2022, AAKBA-EPL2023}. This modified setting allows us to develop models in which the attractiveness of the ground patterns, that are not directly accessible and therefore are not stored in the coupling matrix, emerges as a consequence of the coalescence of attraction basins associated to examples related to the same ground pattern. This kind of phenomenon is responsible for the generalization capabilities of the model. The ways examples can be combined into the coupling matrix mimic the two training protocols: {\it i.} in the \emph{supervised} setting, we {\it a priori} know the organization in classes of the examples, so that -- assuming noise in the examples is uncorrelated -- the empirical mean, say $\bar{\bb \xi}^{\mu}$ for the $\mu$-th class, is a good approximation of the reality; in this way, we can promote this mean as a representative of the given $\mu$-th class, and store this, namely use the set $\{\bar{\bb \xi}^{\mu}\}_{\mu=1,...,P}$ to build $\bb J$; {\it ii.} in the \emph{unsupervised} setting, in which there is no {\it a priori} distinction between examples belonging to different classes, and the only possible way to store information in $\bb J$ is to treat all of them as distinct information patterns. 
\par\medskip
As a matter of fact, crucial properties of these Hopfield-like models are entirely encoded in the structure of the (random) coupling matrix, and, in particular, in its spectral properties, see for example \cite{benedetti2023eigenvector,regularizationdreaming,leonelli2021effective} for recent investigations. Also, the strong relation between random-matrix theory tools and spin-glass models constitutes a long-standing research topic, see for instance \cite{kosterlitz1976spherical,galluccio1998rational,auffinger2013random,zhu2016inverse}, and also
\cite{pennington2017geometry,mai2018random,liao2018dynamics,seddik2020random,zhou2021eigenvalue,couillet2022random,granziol2022random,barbier2023fundamental,LenkaFlorent,LenkaJPA,
AABF-JPA201} for applications in machine learning and information theory. In this paper, we consider a set of Rademacher ground patterns and we obtain a sample of examples by randomly flipping a certain fraction of their entries. With this sample, we build the dreaming kernel, distinguishing between a supervised and an unsupervised version and for both we derive the exact eigenvalue distribution in the limit of a large size $N$. By relying on such a knowledge, we inspect the generalization capabilities of the model, as a function of its parameters.
\par\medskip
The path that we pursue is the following: first, we present the model and the related definitions (Sec.~\ref{sec:framework}), next, we state the main analytical results (Sec.~\ref{sec:results}), and we apply them to investigate the information-processing capabilities of the model (Sec.~\ref{sec:attractive}); finally, we summarize and discuss our findings (Sec.~\ref{sec:conclusions}). The proofs and the technical details are collected in the Appendices.

\section{The framework: models, methods and quantities}\label{sec:framework}
Given a set of patterns $\bb \xi^\mu \in \{-1, +1\}^N$, with $\mu=1,...,K$, the reference coupling matrix for the following analysis is given by 
\begin{equation}
	\label{eq:coupling_matrix}
	J_{ij}^{\bb\xi}(t)=\frac1N \sum_{\mu,\nu=1}^K \xi^\mu_i \Big(\frac{1+t}{\bb 1+t \bb C}\Big)_{\mu\nu}\xi^\nu_j.
\end{equation}
where $t \in \mathbb R^+$ and $\bb C$ is the pattern correlation matrix (\emph{vide infra}). 
This matrix was introduced in \cite{fachechi2019dreaming} and can be derived from Hebb's one by implementing consolidation and remotion mechanisms inspired by those occurring in mammal's brain during sleep. Thus, the resulting model is referred to as ``dreaming Hopfield model'' and $t$, which tunes the extent of such mechanisms, as ``dreaming time'', see also the recent related works \cite{zanin23,serricchio23,camilli23,benedetti2023eigenvector,VCMZ-2023}. The reason why we are focusing on $\bb J^{\bb \xi}$ is that it includes paradigmatic cases: by setting $t=0$ we recover the Hebbian coupling and, in the limit $t \to \infty$, we recover Kohonen's projection matrix \cite{Kohonen-1972}; the latter is known to reach the storage-capacity upper-bound, that is, a number $K=N$ of patterns can be successfully stored and retrieved. Moreover, the coupling matrix \eqref{eq:coupling_matrix}
turns out to emerge as the solution of the minimization of a $L_2$-regularized loss-function where a cost is shaped whenever the configuration corresponding to one of the stored patterns is not stable and where the regularization parameter mapped into the dreaming time \cite{regularizationdreaming}.
In fact, the dreaming time controls the overlap between different attraction basins: the higher $t$, the lower the attractive power of spurious configurations \cite{fachechi2019dreaming,agliari2019dreaming,AlbertIEEE}. 

Before proceeding, it is worth introducing the following notation $x \sim \textrm{Rad}(p)$, with $p \in [-1,+1]$, that, in the following, shall be exploited to define a binary random variable $x$, drawn from the distribution $P(x) = \frac{1-p}{2} \delta_{x,-1} +  \frac{1+p}{2} \delta_{x,+1}$, in such a way that, when $p=0$, $x$ is a standard Rademacher variable, while, when $p\neq0$, $x$ is  a biased binary random variable with expectation $p$. 
We are now ready to describe the three settings that we are inspecting in the next sections:
\begin{itemize}
    \item[a)] In the basic \emph{storing} setting, we have $K=P$ patterns\footnote{In the basic storing settings, we use $P$ to denote the number of orthogonal patterns for homogeneity with other scenarios, in which $P$ in the number of classes, since the relevant parameter will always be $\alpha=P/N$ in the thermodynamic limit.} $\{\bb \xi^\mu\}_{\mu=1}^P$, each made of $N$ Rademacher entries: $\xi_i^{\mu} \sim \textrm{Rad}(0)$ for any $\mu=1,\dots,P$ and $i=1,\dots ,N$. The sum over $\mu, \nu$ \eqref{eq:coupling_matrix} is performed over $\mu,\nu = 1,\dots,P$ and $C_{\mu \nu} = \frac{1}{N}\sum_i \xi_i^{\mu}\xi_i^{\nu}$.
    \item[b)] In the \emph{supervised-storing} setting, we have $P$ patterns $\{\bb \zeta^\mu\}_{\mu=1}^P$, each made of $N$ entries, that are meant as ground patterns. From these, we generate $K = P \times M$ examples, denoted as $\{ \bb \xi^{\mu,A} \}_{\mu=1,...,P}^{A=1,...,M}$, by flipping randomly the entries of the related ground pattern, also referred to as archetype. Specifically, we choose Rademacher ground patterns, that is $\zeta_i^{\mu} \sim \textrm{Rad}(0)$ for any $\mu=1,\dots,P$ and $i=1,\dots ,N$, and uncorrelated noise for examples, that is,
     $$
   \xi^{\mu,A}_i= \chi^{\mu,A}_i \zeta^\mu_i,
    $$
    with $\chi^{\mu,A}_i \sim \textrm{Rad}(r)$, for any $\mu=1,\dots,P$, $i=1,\dots ,N$, and $A=1,\dots,M$. 
    In this supervised setting, since the class of each example is unveiled, we can calculate the empirical mean of examples in each class, i.e.
    $$
   \bar  \xi^\mu_i := \frac1M \sum_{A=1}^M \xi^{\mu,A}_i=\frac1M \sum_{A=1}^M \chi^{\mu,A}_i\zeta^\mu_i =: \bar{\chi}^{\mu}_i\zeta^\mu_i.
    $$
    Then, the coupling matrix is defined as
    $$
    J_{ij}^s(t)=\frac1N \sum_{i,j=1}^N \sum_{\mu,\nu = 1}^P \sigma_i\zeta^\mu_i \bar{\chi}_{i}^{\mu} \Big(\frac{1+t}{\bb 1 + t \bb C^{\text s}}\Big)_{\mu\nu} \bar{\chi}_{j}^{\nu} \zeta^\nu _j  \sigma_j,
    $$
    where
    $$C^s_{\mu \nu} =\frac{1}{N} \sum_{i=1}^N \zeta^\mu_i \bar{\chi}_{i}^{\mu} \bar{\chi}_{i}^{\nu} \zeta^\nu_i$$
    is the correlation matrix of the empirical means of the examples.
    \item[c)] In the \emph{unsupervised} setting, the examples $\{ \bb \xi^{\mu,A} \}_{\mu=1,...,P}^{A=1,...,M}$ are generated precisely as in the previous setting b), but, in this case, there is no preassigned label distinguishing between classes. As a consequence, we store all the examples as information patterns, i.e. in \eqref{eq:coupling_matrix} we replace $\xi^\mu_i$ with $\xi^{\mu,A}_i$ and the sum over $\mu$ is replaced with the sum over the category and the example in each class $(\mu,A)$. The coupling matrix is
    $$
    J_{ij}^u(t) =\frac1{NM} \sum_{\mu,\nu=1}^P \sum_{A,B=1}^M \xi^{\mu ,A}_i \Big(\frac{1+t}{\bb1 + t \bb C^u}\Big)_{(\mu A),(\nu B)}\xi^{\nu,B}_i,
    $$
    where
    $$
    C^u_{(\mu A),(\nu B)}= \frac1{NM} \sum_{i=1}^N \xi^{\mu,A}_i \xi^{\nu,B}_i,
    $$
    is the dataset correlation matrix. 
\end{itemize}

\begin{Rmk}
The parameter $r$ represents the fraction of pixels that are expected to be flipped in any example, say $\bb \xi^{\mu,A}$, with respect to the ground $\bb \zeta^{\mu}$. In particular, when $r=0$, each example is, in the average over the entry-flipping probability, orthogonal to the related archetype, while, when $r=1$, each example is a perfect replica of the related archetype. Thus, $r$ and $M$ can be interpreted as, respectively, a measure of the \emph{quality} and of the \emph{quantity} of the available dataset.
\end{Rmk}

\section{Algebraic properties of the coupling matrices} \label{sec:results}

The retrieval capabilities of the models described in the previous section can be addressed by relying on the eigenvalue distributions of the related coupling matrices\footnote{A derivation of these coupling matrices from statistical inference can be found in \cite{pallara} for the standard Hebbian model and in \cite{lad} for the dreaming Hopfield model}. Also, by comparing their spectra we can assess to what extent the models encoded by $\bb J^s$ and $\bb J^u$ differ from the model built on ground patterns and therefore the effectiveness of the definitions for $\bb J^s$ and $\bb J^u$. This motivates the aim of this section, that is, determining the spectral properties for the models under consideration.

\begin{Def}[Thermodynamic limit]
	The termodynamic limit (TDL) is defined as $N,P\to\infty$ with $P=P(N)$ and $\lim_{N\to\infty}P/N=\alpha$, with $0<\alpha \leq 1$.
This coincides with the so-called high-storage regime of the Hopfield model.
\end{Def}
%
In the following, unless it is explicitly specified, we will denote the coupling matrix as $\bb J(t)$, regardless of the setting under consideration. In fact, by denoting with $\bb X$ the matrix made of the information vectors (patterns or examples) on the rows -- in the random pattern and supervised cases, it is a $P\times  N$ matrix with entries resp. $X_{\mu i}= \xi^\mu_i$ and $X_{\mu i}= \frac1M \sum_A \chi ^{\mu,A}_i \zeta^\mu_i$, while in the unsupervised case it is a $MP \times N$ matrix with entries $X_{(\mu,A),i}=\chi^{\mu,A}_i \zeta^\mu_i$ where the double index $(\mu,A)$ labels each example in the dataset -- any of the coupling matrices introduced above can be written as 
\begin{equation}\label{eq:what}
\bb J(t) = \frac1{D_N} \bb X ^T \Big(\frac{1+t}{\bb 1+t \bb C}\Big)\bb X,
\end{equation}
with $\bb C= \frac1{D_N} \bb X \bb X^T$ and $D_N$ is a normalization factor that reads as $D_N=N$ for the basic storing and the supervised case, while $D_N=NM$ for the unsupervised case. 
\begin{Lem}\label{lem:1}
	The following results hold:
			\begin{enumerate}
		\item The coupling matrix $\bb J(t)$ satisfies the differential equation
		\begin{equation} \label{eq:Jevol}
			\bb {\dot{ J}}(t) = \frac{1}{1+t}[\bb J(t)-\bb J(t)^2].
		\end{equation}
		\item Given $\lambda_\alpha ^0$ the eigenvalues of the coupling matrix $\bb J^0=\bb J(0)$, then the eigenvalues $\lambda_\alpha (t)$ of $\bb J(t)$ are in bijective correspondence with $\lambda_\alpha^0$ through the relation
		\begin{equation}\label{eq:lambda_streaming}
			\lambda_\alpha (t)= \frac{1+t}{1+t\lambda_\alpha^0}\lambda_\alpha^0.
		\end{equation}
	\item The eigenspaces of $\bb J(0)$ are stable under dreaming flow, that is, if $\{v_\alpha^1,\dots, v_\alpha^m\}$ are the eigenvectors of $\bb J^0$ associated to an $m$-degenerate eigenvalue $\lambda_\alpha^0$, then $Span(\{v_\alpha^1,\dots, v_\alpha^m\})$ is the eigenspace of $\bb J(t)$ associated to the eigenvalue $\lambda_\alpha(t)$.	

	\end{enumerate}
\end{Lem}


The proof of this lemma is detailed in App.~\ref{app:A0}.

\begin{Rmk}
	Lemma \ref{lem:1} establishes that the dreaming interaction matrix $\bb J (t)$ defined in \eqref{eq:coupling_matrix} results from the evolution \eqref{eq:Jevol}, regardless of the underlying setting. In other words, whether $\bb J (t)$ stems from a basic storing or by the combination of corrupted examples (either labelled or not), that is, whether $\bb J (t)$ is meant for storing or for generalization, it still results from the process represented by \eqref{eq:Jevol} which encodes for a consolidation and a remotion mechanism.
\end{Rmk}

\par\medskip

Having established these basic properties of the coupling matrices in all the three settings under consideration at finite $N$, $P$ and $M$, we are now able to study their relevant spectral properties in the thermodynamic limit. In that limit, for the supervised and unsupervised settings, we also pose $M\to\infty$, regardless of $N$; this condition is also referred to as the big-data regime. The main results of the Section are summarized in the following

\begin{Thm}\label{thm:1} In the thermodynamic limit $N \to \infty$ and, for the supervised and unsupervised settings, in the infinite sample-size limit $M\to\infty$, the following results hold:
\begin{enumerate}
			\item The empirical spectral distribution $\mu_N^0(\lambda) = \frac1N \sum_\alpha \delta_{\lambda^0_\alpha,\lambda}$ of $\bb J^0$ converges in weak topology $\mu_N ^0\to \mu^0$, where
		\begin{equation}
			d\mu^0(\lambda)= (1-\alpha)\delta(\lambda-\lambda^0_{\text{peak}})d\lambda+\alpha d\mu_{\text{MP}}( \lambda),
		\end{equation}
	with the measure $d\mu_{\text{MP}}(\lambda)$ being a shifted Marchenko-Pastur distribution $\text{MP}(\alpha,\sigma^2)$, i.e.
		\begin{equation}
			d\mu_{\text{MP}}(\lambda)=\frac{1}{2\pi \sigma^2}\frac{\sqrt{(\lambda^0_+-\lambda)(\lambda-\lambda^0_-)}}{\alpha (\lambda-\delta)}d\lambda,
		\end{equation}
		and $\lambda^0_\pm = \sigma^2(1\pm \sqrt \alpha)^2+\delta$. The parameters $\delta$, $\sigma^2$ and $\lambda^0_{\text{peak}}$ depend on the setting under consideration;
		\item
		The empirical spectral distribution $\mu_N^t(\lambda) = \frac1N \sum_\alpha \delta_{\lambda_\alpha(t),\lambda}$ of the coupling matrix $\bb J(t)$ converges in weak topology $\mu_N ^t\to \mu^t$, where
		\begin{equation}
			d\mu^t (\lambda)= d\mu^0 \Big[\frac{\lambda}{1+t(1-\lambda)}\Big],
		\end{equation}
	i.e.
	\begin{equation}
 \label{eq:pdfs}
		d\mu^t (\lambda)=(1-\alpha)\delta\Big[\lambda- \frac{(1+t)\lambda_{\text{peak}}^0}{1+t\lambda_{\text{peak}}^0}\Big]d\lambda+\alpha d\mu^t_{\text{bulk}},
	\end{equation}
where the bulk distribution is
\begin{equation}
	d\mu^t_{\text{bulk}}(\lambda)= \frac{1+t}{2\pi \sigma^2}\frac{\sqrt{(1+t\lambda_-^0 )(1+t\lambda_+^0 )}}{[1+t(1-\lambda)]^2}\frac{\sqrt{(\lambda_+ -\lambda)(\lambda-\lambda_-)}}{\alpha[(1+t\delta)\lambda -{(1+t)\delta}]}d\lambda,
\end{equation}
with
\begin{equation} \label{eq:pm}
	\lambda_\pm = \frac{(1+t)\lambda_\pm ^0}{1+t \lambda_\pm ^0},
\end{equation}
and $\delta$, $\sigma^2$ and $\lambda_{\text{peak}}^0$ are the parameters of previous point depending on the setting under consideration.
\end{enumerate}
\end{Thm}
The proof of this theorem is provided in App.~\ref{app:A}.

\begin{figure}
    \centering
    \includegraphics[width=\textwidth]{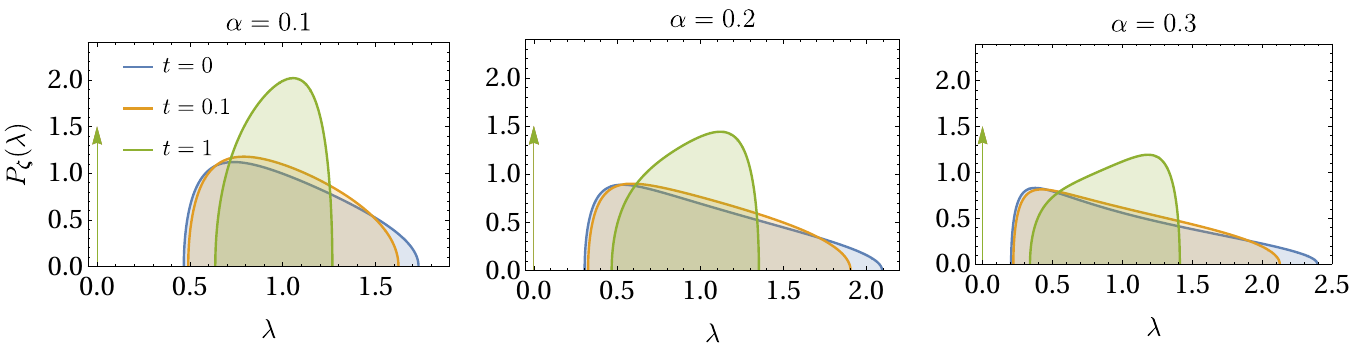}
     \includegraphics[width=\textwidth]{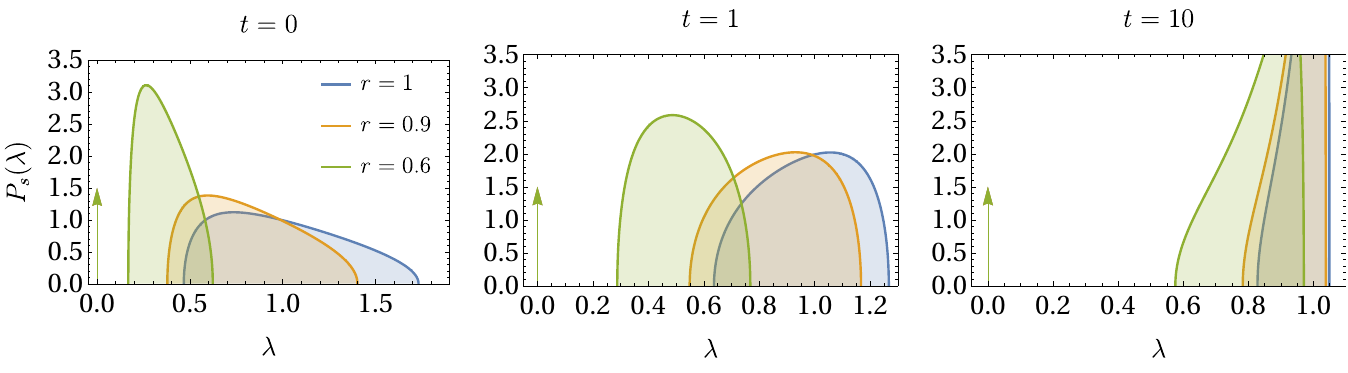}
      \includegraphics[width=\textwidth]{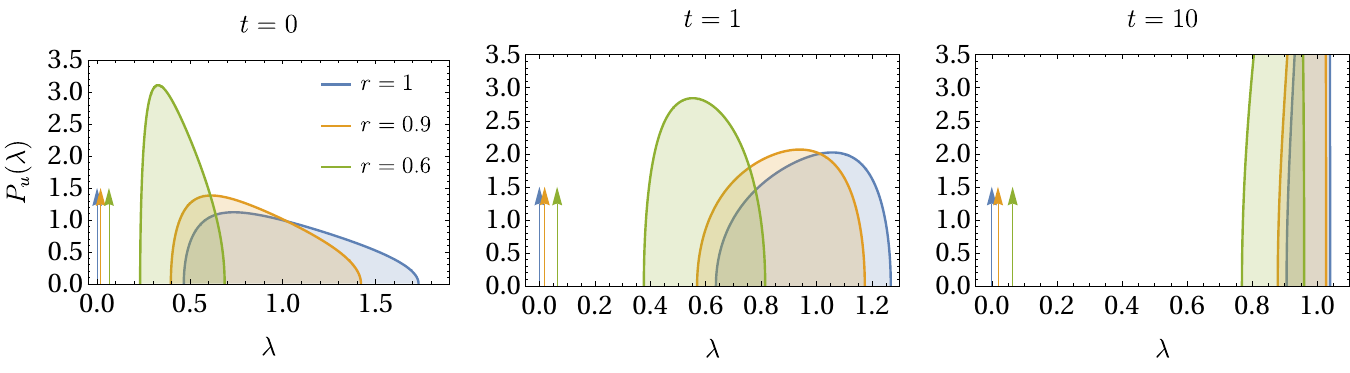}
    \caption{{\bfseries Limiting spectral distributions of the couplings matrix.} The figure shows the probability distribution $P(\lambda)= \frac{d \mu}{d \lambda}$ \eqref{eq:pdfs} in the three settings under consideration: basic storing (first row), supervised (second row) and unsupervised (third row) cases. In the first row, we plotted the spectral distribution for various values of $\alpha$ and $t$, while in the supervised and unsupervised setting we fixed $\alpha=0.1$ and vary $t$ and $r$. The vertical arrows (whose heights are arbitrary) refer to the location of the $\delta$-peak: in the basic storing and supervised cases, the location is at $\lambda=0$ (as $\lambda_{peak} =0$), while in the unsupervised setting it depends on $\alpha$, $t$ and $r$, as foreseen by Thm. \ref{thm:1}.}
    \label{fig:pdfs}
\end{figure}

\begin{Rmk}
At $t=0$ and $r=1$, all the three cases reduces to the usual Marchenko-Pastur theorem \cite{MP}. At $t\to\infty$, the basic storing setting and supervised/unsupervised cases at $r=1$ reproduce the spectrum of the projector model \cite{Kanter-Sompolinsky-1987}, consisting in two $\delta$-peaks at $\lambda=0$ and $\lambda=1$, with mass resp. $1-\alpha$ and $\alpha$.

\end{Rmk}
Thm. \ref{thm:1} fully characterizes the spectral properties of the coupling matrix in all the three settings. These results are used to unveil the role of the dreaming parameter $t$, especially in understanding how empirical scenarios (i.e., supervised and unsupervised settings) deviate from their {\it ideal} counterpart, that is, the coupling matrix $\bb J^{\bb\zeta}$ realized with the hidden ground-truths $\bb\zeta^\mu$. Also, spectral tools will be used to derive results about retrieval properties of these models as associative memories. This will be the subject of the rest of the paper. Before going further, we want to stress that unsupervised setting is a rather peculiar scenario (if compared to basic storing and supervised cases), as it is characterized by two different regimes. If $MP\ge N$, the coupling matrix is full-rank, and eigenvalues are all strictly positive (this is the regime in which Thm. \ref{thm:1} is derived, as we are interested in the large dataset limit). At finite $M$, the spectrum consists in two separate bulks (as can be checked numerically) and for $M\to\infty$ the lowest component converges to a $\delta$-peak located at $\lambda_{{peak}}=\alpha (1-r^2)$ with mass $1-\alpha$. The remaining fraction $\alpha$ of eigenvalues collapses to $\lambda=1$ in the $t\to\infty$ limit (as can be checked by inspecting Eq. \eqref{eq:pm}), which becomes $P$-degenerate and the corresponding eigenspace coinciding with the linear space spanned by the ground-truth $\bb\zeta^\mu$ (see Proof of Thm. \ref{thm:1}). In the $MP<N$ case, the coupling matrix is low-rank, with a fraction $1-MP/N$ of vanishing eigenvalues. In this regime, the positive component of the spectrum exhibits a different distribution, with $K$ large top eigenvalues, well-separated from the continuous bulk at low $t$, which ultimately collapse to $\lambda=1$ in the $t\to\infty$ limit. The emerging phenomenology is totally different in that case and it is beyond the scope of this paper, so we refer to \cite{regularizationdreaming} for a deeper discussion. A visual representation of the spectral distributions in the three settings is provided in Fig. \ref{fig:pdfs}. 



\par\medskip

When dealing with examples of unavailable, ground patterns, either in a supervised or unsupervised setting, it is natural to question whether our empirical models accounts for a good representation of the reality, namely whether $\bb J^s$ and $\bb J^u$ are close to $\bb J^{\bb \zeta}$ where we directly store the ground-truths as information patterns. In order to assess the validity of our models, we consider the squared error between the empirical coupling matrices and the one built with the ground-truths, as the parameter $\alpha$, $r$ and $t$ are tuned.

\begin{Def}
	The Squared Error (SE) between empirical and ground-truth coupling matrices is defined as
	\begin{equation} \label{eq:exdeltaM}
	\delta^{s,u}_M (\alpha,r,t) =\frac1N \lVert   \bb J^{\bb \zeta}(t)-\bb J^{s,u}(t)\lVert_F^2
	\end{equation}
	where the superscripts $s,u$ label the supervised or unsupervised setting, $\bb J^{\bb\zeta}(t)$ is the coupling matrix built with the ground-truths, and $\lVert \cdot \lVert _F$ is the Frobenius norm between matrices. We denote $\delta^{s,u}(\alpha,r,t)=\lim_{M\to\infty} \delta_M^{s,u}(\alpha,r,t)$.
\end{Def}

\begin{Prp} \label{prop:1}
	In the thermodynamic limit, and for $M\to\infty$, the SE can be expressed as
	\begin{equation} \label{eq:exdelta}
		\delta^{s,u}(\alpha,r,t)= \int \big [\lambda - f^{s,u}_{r,t}(\lambda)\big]^2d\mu_{\zeta}^t(\lambda),
	\end{equation}
where $\mu_{\zeta}^t$ is the limiting spectral distribution of $\bb {J^{\zeta}}$, and
	\begin{enumerate}
			\item in the supervised setting:
			$$
			f^s_{r,t}(\lambda)=\frac{\lambda  r^2 (t+1)}{\lambda  \left(r^2-1\right) t+t+1};
			$$
			\item in the unsupervised setting:
			$$
			f^u_{r,t}(\lambda)=\frac{(t+1) \left\{\lambda  r^2+\alpha  \left(r^2-1\right) [(\lambda -1) t-1]\right\}}{\lambda  \left(r^2-1\right)t (\alpha  t+1)-\left[\alpha  \left(r^2-1\right) (t+1) t\right]+t+1}.
			$$
	\end{enumerate}
\end{Prp}

Again, we refer to the Appendices and, specifically to App.~\ref{app:B}, for the complete proof.

\medskip

\begin{figure}[tb]
    \centering
    \includegraphics[width=\textwidth]{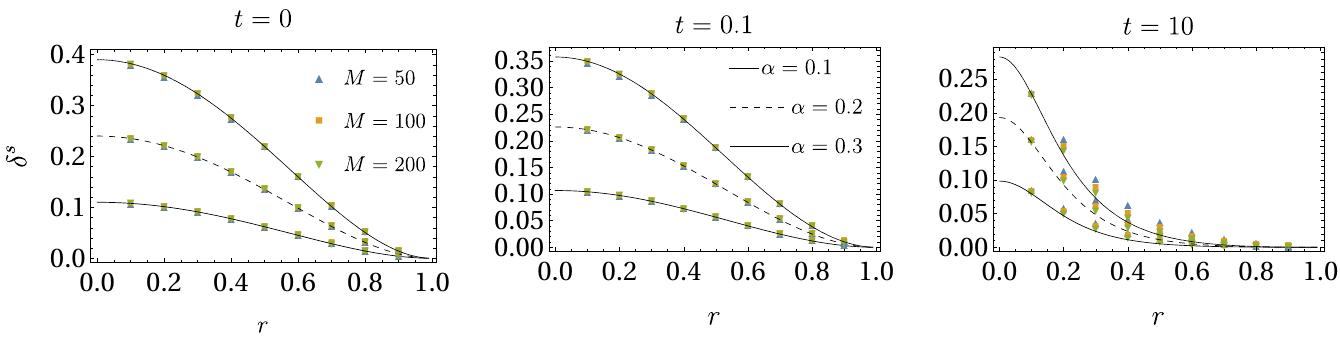}
    \includegraphics[width=\textwidth]{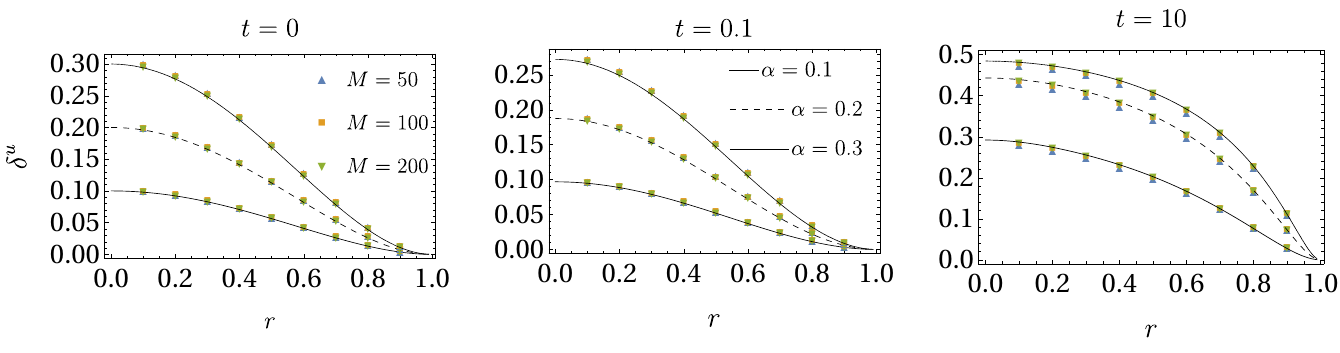}
    \caption{{\bfseries Squared error for supervised and unsupervised settings.} The figure shows the comparison between numerical results for the SE \eqref{eq:exdeltaM} at finite $M$ and the theoretical prediction for $M\to\infty$ in the thermodynamic limit as a function of $r$ for various values of $\alpha$ and $t$. The first row refers to the supervised setting, while the second line shows the results for the unsupervised case. For fixed $t$, each plot exhibits the results for $\alpha=0.1$ (solid black curve), $\alpha=0.2$ (dashed black curve) and $\alpha=0.3$ (dotted black curve), while the markers refer to $M=50,100,200$. The network size is fixed to $N=1000$ in all cases.}
    \label{fig:metrics}
\end{figure}

The exact SE $\delta^{s,u}(\alpha,r,t)$, obtained by evaluating \eqref{eq:exdelta}, is plotted versus $r$ in Fig.~\ref{fig:metrics} and several values of $\alpha$ and $t$ are considered. Also, these theoretical results, valid in the limit $M \to \infty$, are compared with the numerical evaluation of $\delta_M^{s,u}(\alpha,r,t)$, as reported in \eqref{eq:exdeltaM}, at finite sample size $M$. As is clear, there is a perfect agreement of the numerical results with the theoretical predictions, and the accuracy of theoretical results gets better by increasing the dataset size $M$. Trivially, the empirical versions of the coupling matrix do converge to the basic storing setting if $r$ is high enough. However, the interesting point is to analyze the role of the dreaming parameter $t$. In the supervised setting, increasing the dreaming time results in a faster convergence of the coupling matrix towards the basic storing setting (see e.g. the case $t=10$ in the first row of Fig.~\ref{fig:metrics}). 
The reason lies in the fact that dreaming mechanism ensures that stored configurations are dynamically stable. This can be easily understood in the basic storing setting at $t\to\infty$. In that case, the coupling matrix reduces to the projector model \cite{Kanter-Sompolinsky-1987}, for which $\bb J \cdot \bb \xi^\mu=\bb \xi^\mu$: thus, patterns becomes fixed point under the usual neuronal dynamics ({\it vide infra}). Equivalently, one can interpret the projector model $\frac1N \bb\xi^T \bb C^{-1}\bb \xi$ as a Hebbian prescription for the rotated vectors $(\bb C^{-1/2}\bb \xi)_\mu$: in this case, the stored configurations are forced to be orthogonal, and spurious correlation between patterns (which, in the usual setting of the Hopfield model, are responsible for the breakdown of retrieval capabilities above the critical storage capacity \cite{Amit}) is filtered out. In the unsupervised setting, the dreaming mechanism works in the same way but this time the patterns that are stored are the single examples rather than their empirical mean. In this setup, correlation between examples within the same class is crucial and should be preserved, as it allows for generalization purposes, see \cite{regularizationdreaming}. Thus, as $t$ gets larger, provided that the overall number of examples is not too large, each single example can be a fixed point. This way, increasing $t$ too much would prevent the convergence of $\bb J^u$ to $\bb {J^{\zeta}}$, unless the examples are perfect realizations of the ground-truths, i.e. $r=1$.

\section{Spectral tools at work: an application to retrieval}\label{sec:attractive}

The Hopfield model and its variations are nothing but spin-glasses with a Hebbian-like prescription for the interactions, and the structure of the quenched disorder encoded in the coupling matrix is known to govern the thermodynamic behavior of the statistical-mechanical model, see for instance \cite{sommers-PRL1988,galluccio-physa1998,
kanaka-PRL2006,castillo-PRE2008,AABF-JPA2019}. Thus, it is reasonable to expect that crucial properties of Hopfield-like models strongly depend on the spectral details of the interaction matrix $\bb J$. In this Section, we aim to provide details about the functioning of these models by simply applying the results derived so far. Let us start from the (deterministic) parallel dynamics, which reads as
\begin{equation}
	\label{eq:neural_dynamics}
	\bb \sigma^{(n+1)}=\sgn [\bb J(t)\cdot\bb \sigma^{(n)}].
\end{equation}
We stress that, here, the evolution time is represented by the integer $n$, while $t$ is the dreaming time that is retained fixed: synaptic weights are quenched during the neural dynamics.  
We are primarily interested in the stability of specific configurations, that is, in the probability that the system in a configuration $\bb\sigma^{(0)}$ at time $n$ will be in the same configuration at time $n+1$. 
In order to analyze the stability of a given initial configuration $\bb\sigma^{(0)}$, we consider the 1-step update of the neural network:
\begin{equation}
	\label{eq:fixed_point}
	\sigma_i ^{(1)} = \sgn \Big [ \sum_{j=1}^N   J_{ij}(t) \sigma_j^{(0)} \Big ].
\end{equation}
We say that the configuration $\bb\sigma^{(0)}$ is stable at the neuron-index $i$ if $\sigma^{(1)}_i = \sigma_i^{(0)}$. Although considering the neural configuration after 1-step evolution could appear a rather limited problem, as we will see, it is enough to understand important properties about the evolution of the model under consideration. Also, we incidentally notice that this is a standard time span in machine-learning training algorithms like CD-1 \cite{CD} and that checking the stability in the 1-step dynamics can  be recast in checking the stability by signal-to-noise-techniques \cite{AAKBA-EPL2023}.
\newline
By multiplying both sides of eq.~\eqref{eq:fixed_point} by $\sigma_i ^{(0)}$ and by exploiting the binary nature of $\sigma_i^{(0)}$, we get
$$
\sigma_i ^{(1)} \sigma_i^{(0)} = \sgn \Big [\sum_{j=1}^N   J_{ij}(t)  \sigma_j^{(0)}  \sigma_i^{(0)}\Big],
$$
which is a key quantity for studying the stability of $\sigma_i ^{(0)}$: if the argument of the sign function is positive (negative), the spin $i$ is stable (unstable). In fact, $\frac{1}{2}[1 - \frac{1}{N} \sum_i \sigma_i^{(1)}\sigma_i^{(0)}]$ represents the fraction of neurons that change their state in the first step of the dynamics, namely the Hamming distance between $\bb \sigma^{(0)}$ and $\bb \sigma^{(1)}$. Thus, we introduce the following
\begin{Def}
    Given a configuration ${\bb\sigma}$, the {\it stability} of its $i$-th neuron is
    \begin{equation}
	\label{eq:stability}
	 \Delta_i ({\bb\sigma}) = \sum_j J_{ij}(t)   \sigma_j \sigma_i.
\end{equation}
\end{Def}
This nomenclature is adopted to get close to the dictionary used in \cite{benedetti2023eigenvector}.
More generally, one could be also interested in measuring the overlap between a specific state $\bb x$ and the configuration of the network after a single update step starting from a reference configuration ${\bb\sigma}^{(0)}$. This quantity would be related to the capability of the configuration $\bb x$ to ``attract'' the dynamics when preparing the system in $\bb{\sigma}^{(0)}$, since we are measuring the overlap with $\bb x$ after the update. Thus, we are also interested in the computation of the following quantity:
\begin{Def}
Given two configurations $\bb\sigma$ and $\bb x$, the {\it attractiveness} of $\bb x$ w.r.t. ${\bb\sigma}$ is
    \begin{equation}
    \label{eq:attractiveness1}
    \Delta_i (\bb x, {\bb\sigma})=\sum_j J_{ij}(t)  \sigma_j x_i.
\end{equation}
More generally, the attractiveness of $\bb x$ w.r.t. a set $\Omega\in \Sigma_N$ is 
\begin{equation}
\label{eq:attractiveness2}
    \Delta_i (\bb x, \Omega) = \underset{\bb\sigma \in \Omega}{\text{inf}}\, \Delta_i (\bb x,\bb \sigma).
\end{equation}
Trivially, $\Delta_i ({\bb\sigma},{\bb\sigma}) =\Delta_i ({\bb\sigma})$.
\end{Def}
\begin{Rmk}
    If $\Delta_i({\bb\sigma},{\bb\sigma}) = \Delta_i ({\bb\sigma})\ge 0$ for all $i$, then $\bb \sigma$ is a fixed point for the dynamics. Analogously, if $\Delta_i (\bb x,{\bb\sigma})\ge 0$, then the network evolves from $\bb \sigma$ towards $\bb x$. In other words, the stability and the attractiveness measure the correlation of the network configuration after a single update -- starting from ${\bb\sigma}$ -- w.r.t. ${\bb\sigma}$ itself and  w.r.t. the configuration $\bb x$. 
    \end{Rmk}
    Introducing the configuration overlaps \begin{eqnarray}
    m^{(0)}(\bb x) &=&  \frac{1}{N}\sum_{i=1}^N x_i \sigma_i^{(0)}\\
m^{(1)}(\bb x) &=&\frac{1}{N} \sum_i x_i \sigma_i ^{(1)} = \frac{1}{N} \sum_{i=1}^N \textrm{sgn} [\Delta_i(\bb x, \bb \sigma^{(0)})],
\end{eqnarray}
the vector $\bb x$ is attracting ${\bb\sigma}^{(0)}$ if $m^{(1)}>m^{(0)}$. 
When the configuration $\bb x$ coincides with a pattern, the overlaps above are also known as Mattis magnetizations (related to that pattern) evaluated at time steps $n=0,1$.

The stability and the attractiveness (which we denote in general as $\Delta_i$, as the considered quantity would be clear from the context) defined above are simple tools, but quite natural to understand the retrieval capabilities of attractor neural networks, and are general enough to be handled in any scenario we are interested in. 
In order to simplify the computations and derive clear expressions of the one-step Mattis magnetization in terms of integrals w.r.t. the Marchenko-Pastur distributions, we will make our computations under the following 
\begin{Wrk}[Gaussian approximation]
	Within the Gaussian approximation (GA), we assume that the quantities $\Delta_i$ are
	\begin{itemize}
		\item i.i.d. random variables;
		\item Gaussian distributed.
	\end{itemize}
\end{Wrk}
Clearly, this is not valid in general, however, as we shall see, this assumption -- in all the settings under examination -- leads to a good approximation of the numerical results (see also App.~\ref{app:D}). Thus, within the GA and in the TDL, the 1-step Mattis magnetization w.r.t. the reference configuration $\bb x$ is given by
\begin{equation}
	\label{eq:general_m1}
	m^{(1)}(\bb x)= \frac1N \sum_i \sgn \Delta_i \underset{TDL-GA}\to2P (\Delta \ge 0)-1= \erf\Big[{\frac{\mu_1}{\sqrt{2(\mu_2-\mu_1^2)}}}\Big],
\end{equation}
where $\mu_1$ and $\mu_2$ are the first and (non-centered) second moments for $\Delta$, that is, following the GA, $\Delta_i \sim \mathcal N (\mu_1, \mu_2 -\mu_1^2)$; as we will show in the following $\mu_{1,2}$ can be expressed in terms of integrals over the Marchenko-Pastur law. Within the basic storing setting, the reference configuration will always be a pattern $\bb \xi^\mu$ (either for the stability or the attractiveness), in the supervised/unsupervised cases it will a be a ground-truth $\bb\zeta^\mu$.
To avoid confusions, from now on we will denote with $\mu_{\zeta}^t(\lambda)$ the limiting spectral distribution of the coupling matrix $\bb J^{\bb \zeta}$, while with $\mu^t_s (\lambda)$ and $\mu^t_u (\lambda)$ resp. those in the supervised and unsupervised setting $\bb J^{s}$ and $\bb J^{u}$.

\begin{Rmk}\label{rem:noisyinit}
	The explicit expression of the attractiveness given by Eq. \eqref{eq:attractiveness2} is, in general, out of reach, even within the GA. However, in practice, one is not interested in estimating the attractiveness of a given configuration $\bb x$ w.r.t. an arbitrary set $\Omega$, rather, one is interested in the attractiveness exerted by specific configurations (patterns in the basic storing setting or ground-truths in the supervised/unsupervised ones), that is by $\bb x = \bb \xi^\mu$ or $\bb x =\bb\zeta^\mu$, on states within a certain distance from $\bb x$ itself, as this provides information on the width of their attraction basin. Thus, a natural choice of the sets $\Omega$ is given by the balls $\mathcal B_R (\bb x)$ centered in $\bb x$ of radius $R$, the topology of the balls in $\Sigma_N \equiv \{-1, +1\}^N$ being defined according to the Hamming distance
		$$
		d_H (\bb\sigma,\bb\sigma') = \frac14 \sum_{i=1}^N (\sigma_i - \sigma'_i)^2,
		$$
  being $\bb \sigma$ and $\bb \sigma'$ two configurations in $\Sigma_N$. With this choice, we have that the attractiveness is a function of $\bb x$ and $R$ only, and
	$$
	\Delta_i (\bb x, R) = \Delta_i (\bb x, \mathcal B_R (\bb x))=\underset{\bb\sigma \in \mathcal B_R (\bb x)}{\textrm{inf}}\, \Delta_i (\bb x,\bb \sigma)\approx\underset{\bb\sigma \in\partial \mathcal B_R (\bb x)}{\textrm{min}}\, \Delta_i (\bb x ,\bb \sigma),
	$$
	where the last relation is due to the fact that the least $\bb x$-attracted points are expected to lie on the boundary $\partial B_{R}(\bb x)$. In the large $N$ limit, these boundaries can be realized by perturbing $\bb x$ as $x_i \to x'_i = \eta_i x_i$, with $\eta_i \sim \textrm{Rad}(p)$, in such a way that in the large $N$ limit we have $R(p)=d_H({\bb x'},\bb x) = \frac12\sum_i (1-\eta_i)\approx\frac N2(1-p)$. This motivates why we define the attractiveness of the configuration $\bb x$ as
	\begin{equation}
		\label{eq:pattern_attract_mod}
		\Delta_i (\bb x, R)=\sum_j J_{ij}(t) x_j \eta_j x_i .
	\end{equation}
Clearly, at large $N$ we have $m^{(0)}(\bb x)=p$.
\end{Rmk}


Let us first focus on the basic storing case. In this setting, we are both interested in the stability and attractiveness of the patterns (clearly, at $p=1$, attractiveness and stability coincide). Then, the following proposition holds:
\begin{Prp}\label{prop:pattern_computations}
	In the basic storing setting, within the GA and in the thermodynamic limit:
	\begin{enumerate}
		\item	The empirical first and second moments of the pattern stability are resp.
			\begin{eqnarray}
			\mu_1 &=& \frac1\alpha \int \frac{\lambda^2}{1+t(1-\lambda)}d\mu^t(\lambda) ,\label{eq:patt_c1_stab}\\
			\mu_2 &=& \frac1\alpha \int \frac{\lambda^3}{1+t(1-\lambda)}d\mu^t(\lambda).\label{eq:patt_c2_stab}
		\end{eqnarray}	
	\item The empirical first and second moments of the pattern attractiveness are resp.
		\begin{eqnarray}
		\mu_1 &=& \frac{p}{\alpha}\int \frac{\lambda^2}{1+t(1-\lambda)}d\mu^t(\lambda) ,\label{eq:patt_c1_attr}\\
		\mu_2 &=&({1-p^2})\int \lambda^2 d\mu^t(\lambda)
		+ \frac{p^2}{\alpha}\int \frac{\lambda^3}{1+t(1-\lambda)}d\mu^t(\lambda)  .\label{eq:patt_c2_attr}
	\end{eqnarray}	
	\end{enumerate}
\end{Prp}


The proof of this proposition is provided in App.~\ref{app:D} along with details on the validity of the GA.

\begin{figure}[h!]
\begin{minipage}{0.49\textwidth}
\centering
	\includegraphics[width=\textwidth]{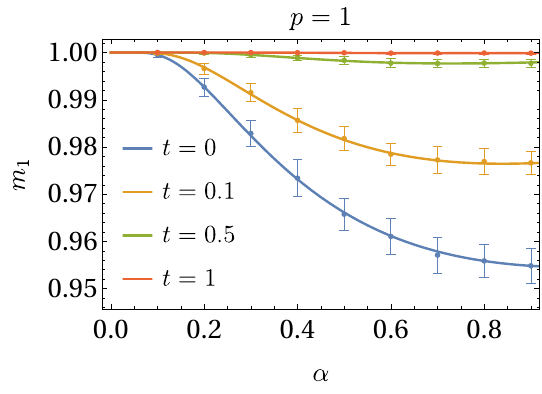}
\end{minipage}
\hfill
\begin{minipage}{0.49\textwidth}
\centering
	\includegraphics[width=\textwidth]{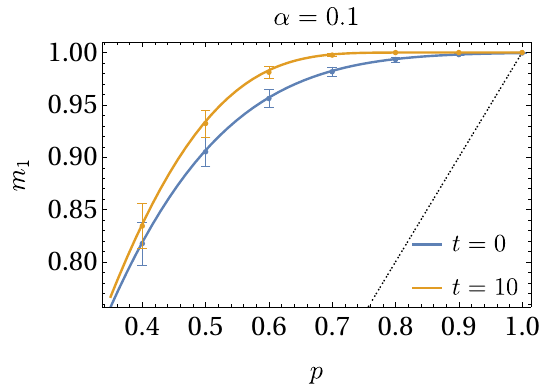}
\end{minipage}
\begin{minipage}{0.49\textwidth}
\centering
	\includegraphics[width=\textwidth]{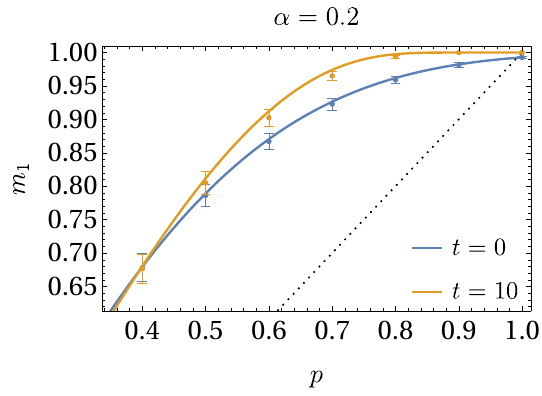}
\end{minipage}\hfill
\begin{minipage}{0.49\textwidth}
\centering
	\includegraphics[width=\textwidth]{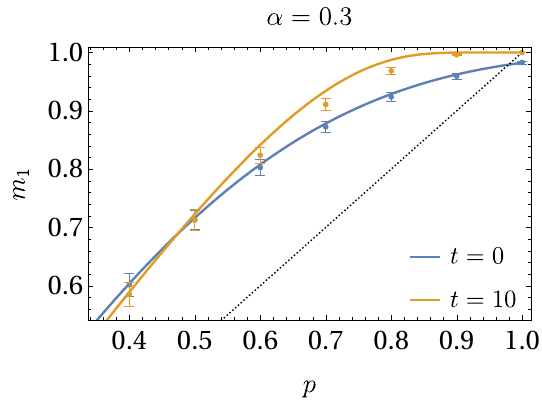}
\end{minipage}
\caption{{\bfseries Stability and attractiveness of patterns in the basic storing setting.} The figure shows a comparison between the theoretical predictions of stability (upper left plot) and attractiveness (other plots). In the former case, the 1-step magnetization $m_1$ starting from one of the patterns ($p=1$) is given as a function of $\alpha$, while for the attractiveness we fixed $\alpha=0.1,0.2,0.3$ and $t=0,10$ (resp. Hebbian and large dreaming time limit) and consider {$m_1$} as a function of the noise level $p$ of the starting configuration, as explained in Rem. \ref{rem:noisyinit}. In these plots, the dashed lines is the identity function $m_1(p)=m_0(p)=p.$ In the numerical simulations, we averaged over 100 different realizations of the patterns for systems with fixed size $N=5000$. In these plots, $m_1$ stands for $m^{(1)}.$}\label{fig:patterns}
\end{figure}
%
%
Once that first two moments are estimated, we can predict the one-step Mattis magnetization according to Eq. \eqref{eq:general_m1}. In particular, at $t=0$, the results reported in Prop. \ref{prop:pattern_computations} lead to $\mu_1=1+\alpha$ and $\mu_2 = \alpha^2+3\alpha+1$, so that
$$
m^{(1)}= \erf\Big(\frac{1+\alpha}{\sqrt{2\alpha}}\Big),
$$
which recovers the well-known expression for the expected magnetization in the Hopfield model \cite{Amit}.\footnote{The factor $1+\alpha$ at the numerator in the error function is due to the fact that we are also including self-interactions. If $J_{ii}=0$, instead, we would have $\erf(1/\sqrt{2\alpha})$ \cite{Rocchi_2017}.} At $t\gg 1$, we get
\begin{eqnarray}
    \mu_1&=& 1-\frac{\alpha }{(\alpha -1) t^2}+\mathcal O(t^{-3}),\\
    \mu_2 &=& 1-\frac{3 \alpha }{(\alpha -1) t^2}+\mathcal O (t^{-3}).
\end{eqnarray}
It follows that
$$
m^{(1)} \underset{t\gg 1} = \erf\Big(\frac{\sqrt{1-\alpha}t}{\sqrt{2\alpha}}
 +\mathcal O (t^0)\Big)\approx 1-\frac1t \frac{\exp(-\frac{1-\alpha}{2\alpha}t^2)}{\sqrt{\pi \frac{1-\alpha}{2\alpha }}}.
$$
Similar results can also be carried out for the pattern attractiveness. The theoretical predictions for the pattern stability and attractiveness, and the relative comparison with numerical results can be respectively found in Fig. \ref{fig:patterns} for different values of the tuneable parameters $\alpha$, $p$, $t$.

As expected, the stability of a pattern is impaired by $\alpha$, but dreaming can mitigate this effect (see the upper left panel in Fig. \ref{fig:patterns}). Dreaming can also enhance the attractiveness of a pattern, yielding a large overlap $m_1 \approx 1$ for a relatively large range of noise values (see the upper right and lower panels in Fig. \ref{fig:patterns}).
\par\medskip
In the supervised and unsupervised settings, the computation of stability and attractiveness of stored vectors follows an analogous route. However, in these cases, rather than in the attracting power of stored (training) examples, we are mainly interested in the {\it generalization} capabilities of the model, that is -- given a starting initial condition with the same statistics of the stored examples -- the attractiveness of ground-truths underlying the training dataset. Thus, in these settings we consider a starting {\it testing} configuration $\bar{\bb\sigma} = \bb \chi \odot \bb\zeta^\mu$ for some $\mu=1,\dots,P$ and $\chi_i^{\mu} \sim \textrm{Rad}(r)$ for any $i, \mu$.\footnote{We stress that validation configurations follows the same statistics as the training examples, but these two sets are independent.} Then, we are interested in the probability that -- after 1-step update -- the neural configuration is aligned with the ground-truth $\bb\zeta^\mu$ generating the testing example. Thus, our major concern in settings $a)$ and $b)$ is the attractiveness of $\bb\zeta^\mu$ w.r.t. testing examples, i.e.
\begin{equation}\label{eq:examples_attract}
	\Delta_i (\bb\zeta^\mu, R(r))=\sum_j J_{ij}(t)\zeta_i^\mu \bar \sigma_j\equiv \sum_j J_{ij}(t)\chi_j \zeta^\mu_j \zeta_i^\mu,
\end{equation}
with such configuration with a distance $R(r)=N(1-r)/2$ from $\bb\zeta^\mu$. Then, the following Proposition holds.

\begin{Prp}\label{prop:un_sup}
	Under the GA and in the thermodynamic limit, the empirical first and second moments of the attractiveness \eqref{eq:examples_attract} read
	\begin{enumerate}
		\item in the supervised setting:
		\begin{eqnarray}
			\mu_1 &=&		\frac1{\alpha r}\int \frac{\lambda^2}{1+t(1-\lambda)} d\mu^t_s(\lambda)	,\\
			\mu_2 &=&	(1-r^2) \int {\lambda^2}d\mu^t_s(\lambda)		+\frac1\alpha\int \frac{\lambda^3}{1+t(1-\lambda)} d\mu^t_s(\lambda)		;
		\end{eqnarray}
	\item in the unsupervised setting:
 \begin{eqnarray}
			\mu_1 &=&		\frac1{\alpha r}\int \frac{\lambda^2}{1+t(1-\lambda)} d\mu^t_u(\lambda)-\frac{1-r^2}{r}\int \lambda d\mu^t_u(\lambda)	,\\
			\mu_2 &=&	\frac1\alpha \int \frac{\lambda^3}{1+t(1-\lambda)} d\mu^t_u(\lambda)		,
		\end{eqnarray}
	\end{enumerate}
\end{Prp}
The proof of this proposition can be found in App.~\ref{app:E}.

\begin{figure}[h!]
    \centering
    \includegraphics[width=\textwidth]{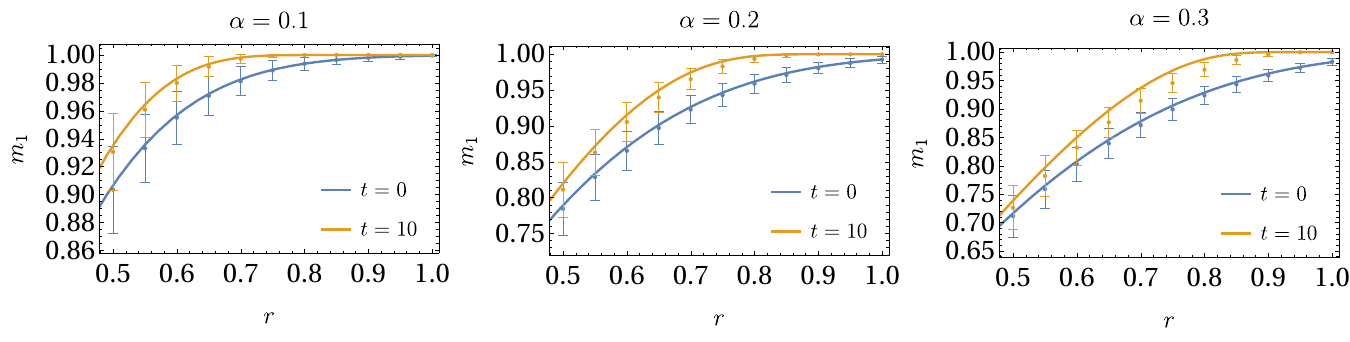}
    \includegraphics[width=\textwidth]{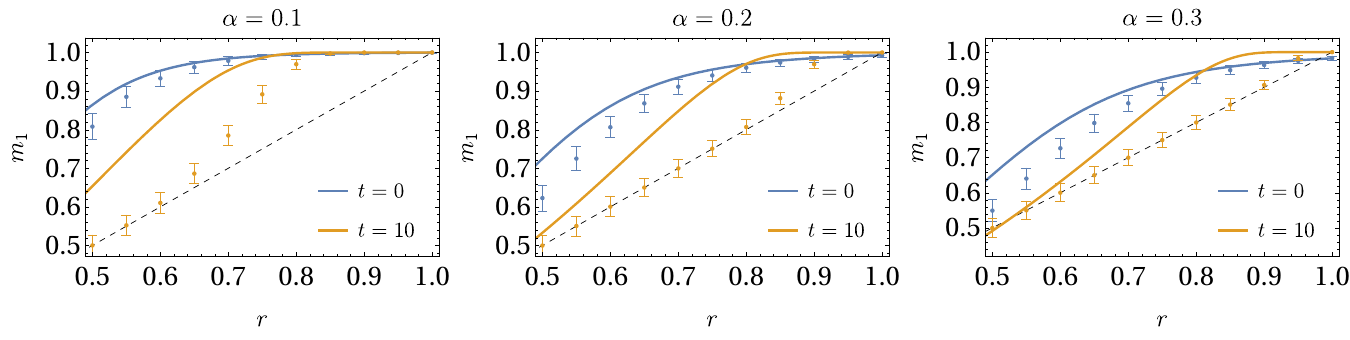}
    \caption{{\bfseries Attractiveness of ground-truths in the supervised and unsupervised settings.} The plot shows a comparison between the theoretical predictions (given by Prop. \ref{prop:un_sup}) of the attractiveness of the ground truths and the numerical results for the supervised (first row) and unsupervised (second row) settings. Numerical results are averaged over 100 different realization of $M=1000$ examples by varying $\alpha$ and $r$, and 100 different realization of the initial conditions. In this case, initial conditions are testing examples, i.e. examples with the same satistics as the the training points, but which are not stored as fixed points. The system size is fixed to $N=1000.$ In the plots, $m_1$ stands for $m^{(1)}.$}
    \label{fig:un_sup}
\end{figure}

In Fig. \ref{fig:un_sup} we show a comparison between the theoretical predictions, as given by Prop. \ref{prop:un_sup}, and numerical simulations. In this case, we used as initial conditions examples with the same satistics as the training points, i.e. with the same quality $r$ w.r.t. the hidden ground patterns. In particular, in the supervised scenario we highlight a positive role of dreaming for any load $\alpha$, while in the unsupervised scenario the effects of dreaming do not have an obvious outcome. In fact, at relatively low load, a large $t$ is detrimental for the retrieval of ground patterns; on the other hand, when $\alpha$ is relatively large, increasing $t$ can be slightly beneficial, at least as long as the dataset is not too corrupted.

\section{Discussion of the results}\label{sec:conclusions}
To conclude this work, we comment the application of our results to the estimation of (1-step) retrieval capabilities of the Hebbian-like models in the settings under examination.
\par
\begin{figure}[h!]
	\centering

	\begin{minipage}{0.47\textwidth}
		\includegraphics[width=\textwidth]{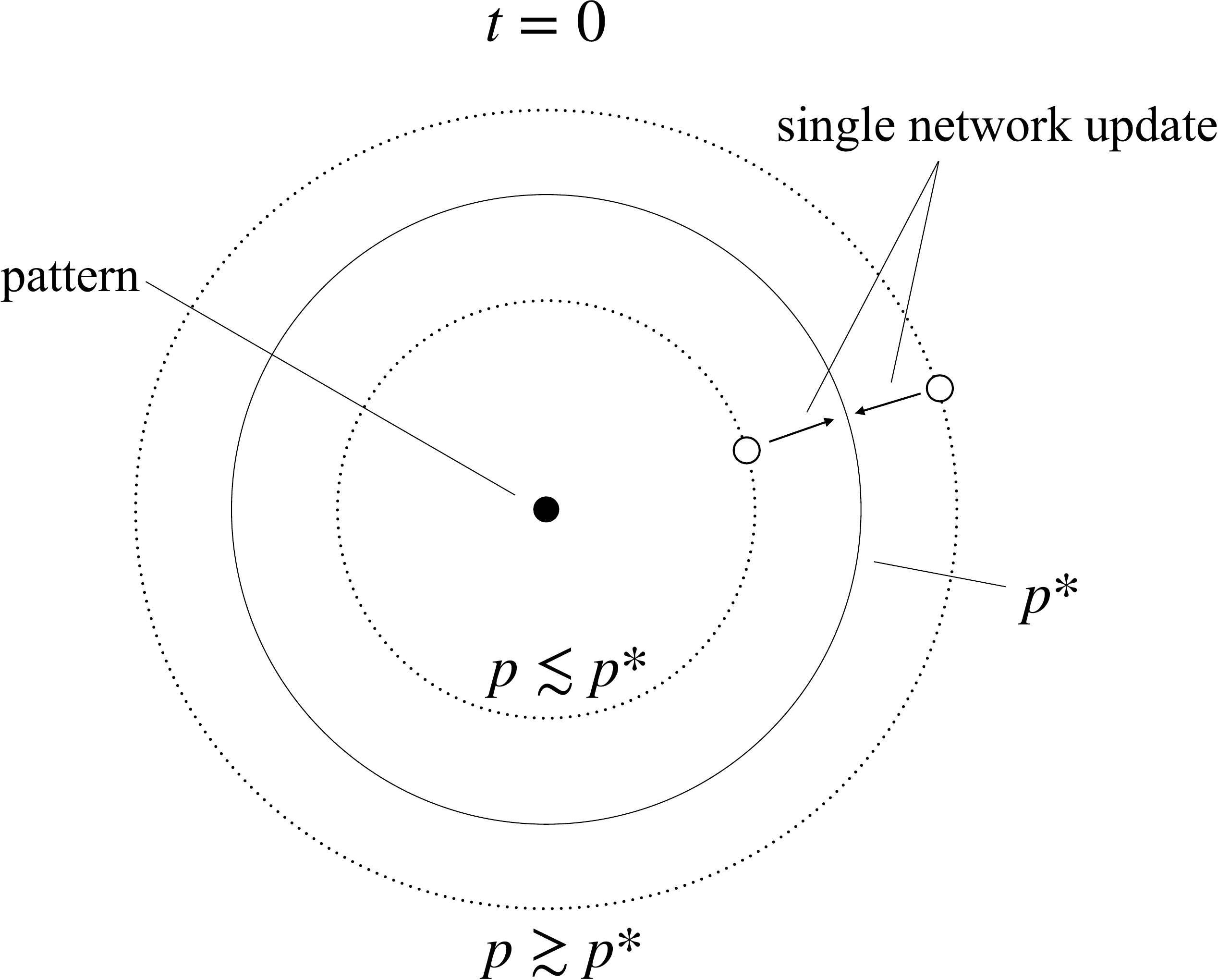}
	\end{minipage}
	\begin{minipage}{0.47\textwidth}
		\includegraphics[width=\textwidth]{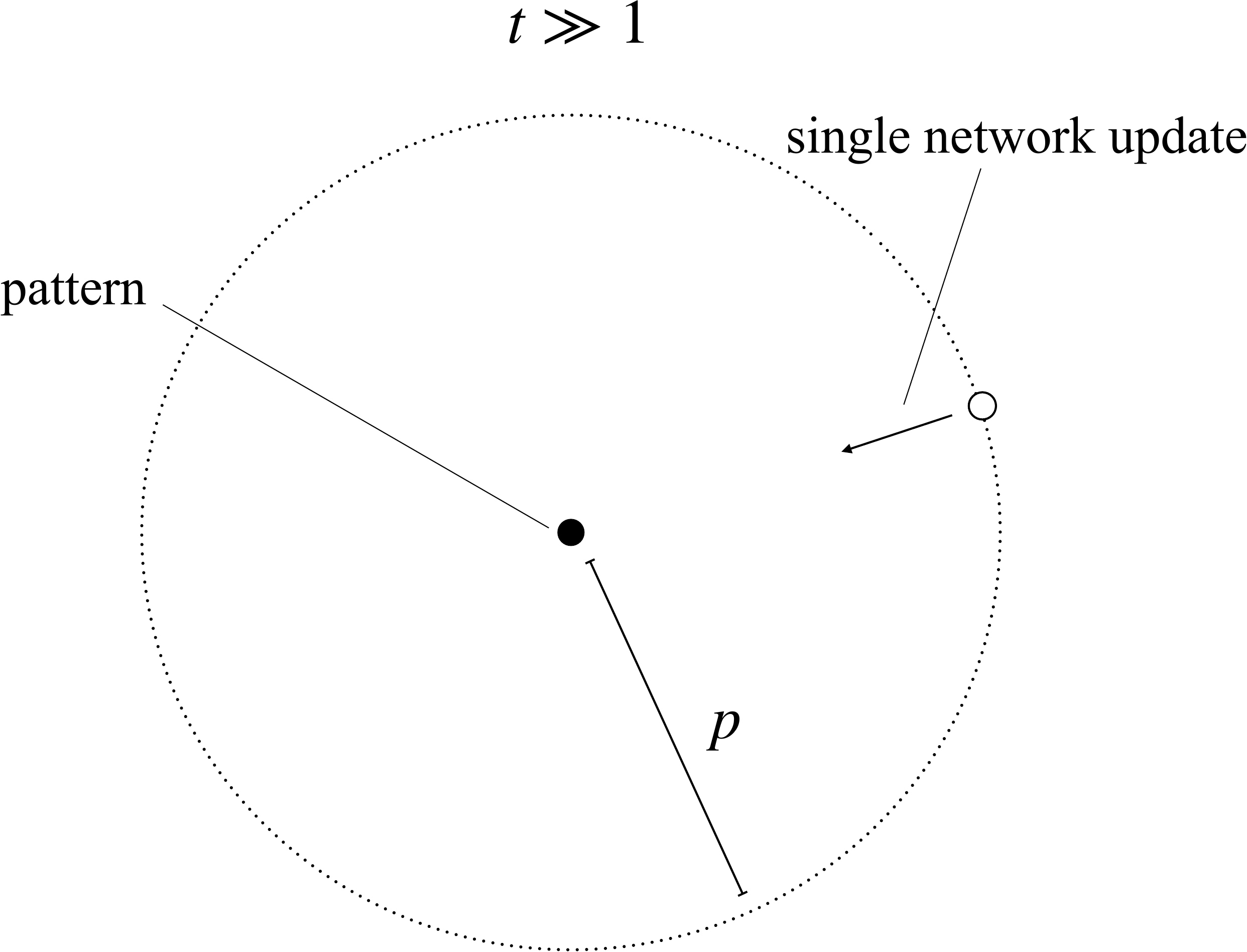}
	\end{minipage}
	\caption{{\bfseries Schematic representation of attractors in the basic storing setting.} The figure shows a pictorial representation of fixed points in the Hopfield model (left) and dreaming model (right) at large dreaming time $t\gg1$. For large $\alpha$ (above the critical storace capacity $\alpha_c=0.14$ for the Hopfield model), fixed points are the balls centered in the pattern with Hamming radius $R(p^*) =\frac N2(1-p^*)$, while in the dreaming model (for large but low enough $t$) patterns are fixed points. At $t\to\infty$, in the dreaming model at $\alpha\le 1$ patterns are always stable configurations for the neural dynamics.}
	\label{fig:detail}
\end{figure}

\medskip
First, let us consider the attractiveness of patterns in the basic storing setting, so we refer to Fig. \ref{fig:patterns}. Looking at the upper right plot ($\alpha=0.1$, just below the critical storage capacity of the Hopfield model $\alpha _c \sim 0.14$), we see that in both cases $t=0$ (Hopfield model) and $t=10$ (essentially the projector case), stored patterns exhibit a strong attractiveness w.r.t. noisy initial conditions. In particular, for all values of $p$, the 1-step magnetization $m^{(1)}$ is always higher than the overlap of the initial condition w.r.t. the considered pattern (the dashed black curve, corresponding to $m^{(1)}(p)=m^{(0)}=p$). The only difference between the two extreme values of the dreaming time consists in the range of $p$ where the one-step update leads to $m^{(1)} \approx 1$: indeed, the dreaming kernel with $t\gg1$ is by far more robust w.r.t. noise in the initial condition. However, as $\alpha$ is increased (above the Hopfield model critical storage capacity, see e.g. $\alpha=0.2$ and $\alpha=0.3$), the situation is different. Indeed, in the $t=0$ case, a non-trivial solution of the equation $m^{(1)}=p$ does appear (this is more evident in the $\alpha=0.3$ case), corresponding to a specific value of the noise in the initial condition, say $p^*$, for which the network update does not lead to a higher magnetization: the network is stacked on the ball $\mathcal B_R (\bb\xi^\mu)$ centered in the pattern $\bb\xi^\mu$ with radius $R(p^*)=\frac N2 (1-p^*)$. Further, for $p> p^*$, after the network update, the final Mattis magnetization is {\it lower} than $p$, meaning that the system is getting farther from the pattern, while for $p<p^*$, the magnetization $m_1$ increases. Even if this is a one-step result, this strongly suggests that, in the Hopfield model at high load, the patterns are no longer stable configurations under neural dynamics, while the fixed points are on the boundary of balls with a non-zero radius ($R(p^*)$ in the one-step case), see the left picture in Fig. \ref{fig:detail} for a schematic representation. This is in agreement with the results reported in \cite{mceliece1987capacity}. Increasing the dreaming time $t$ would heal this behavior: indeed, even at large $\alpha$ the dreaming model will always have $m^{(1)}(p)>p$, and a relatively wide range of $p$ for which $m^{(1)} \approx 1$. Thus, the dreaming model in the basic storing setting always exhibits a better retrieval capabilities w.r.t. the Hopfield model, and -- even for finite $t$ -- the patterns are always attractive, see the image on the right in Fig. \ref{fig:detail}. We stress that the deviation of the numerical results w.r.t. the theoretical predictions has to be ascribed to the gradual break-down of GA.
\par\medskip
The supervised setting shares the same features of the basic storing case, as the empirical mean within classes is -- for dataset with multiplicative noise -- a good prototype of the corresponding ground-truths. In this case, dreaming mechanism would lead in the suppression of fluctuations of this prototype w.r.t. the feature vector $\bb\zeta^\mu$, as we already noticed in Fig. \ref{fig:metrics}. Then, the considerations about attractiveness of patterns in the basic storing setting do hold also in this case, as can be checked by the first row in Fig. \ref{fig:un_sup}.\\
In the unsupervised setting, however, the situation is quite different. First of all, as can be checked in the second row of Fig. \ref{fig:un_sup}, the theoretical predictions exhibit large deviation w.r.t. the numerical results; in particular, the results derived with spectral tools always overestimated the numerical results. This signals a break-down in the statistical independence of terms in the attractiveness, so the GA gets weaker. Despite this deviation, the spectral tools capture the qualitative behavior of the setting. First, we see that, for a high enough noise level in the initial condition (i.e. for relatively small $p$), Hopfield model performs better w.r.t. the large dreaming time limit ($t=10$): this can be seen both from the theoretical predictions and the numerical results. The origin of this opposite behavior can be identified with the fact that, in the unsupervised setting, intra-class correlation is crucial for inferring the ground-truths (i.e. realizing them as mixed linear combinations of stored examples). At low $t$, the attraction basins associated to training points are wide enough to merge an form a single well leading the system to generalize the ground-truths as these mixed configurations are strongly attractive, as clearly explained in \cite{regularizationdreaming}. Increasing $t$ would result in a shrinkage of attraction basins associated to the training points, especially if $r$ is low, so that they are perfectly separated: the system no longer generalize, and we only retrieve training examples (as can be seen from the fact that, at low $r$, the magnetization $m_1$ settles on the identity line $m_1(p)=p$). As a result, in this case training points are stable, i.e. they have large overlap with the eigenvectors of the coupling matrix: this clearly breaks statistical independence needed for GA. Increasing $\alpha$, the Hopfield model undergoes the same behavior of the basic storing and supervised settings (i.e. patterns are no longer fixed points), even in the $r=1$ case (when training points are perfect realizations of the ground-truths), since stored vectors are no longer fixed points, while the dreaming time lead to appreciable results only if the quality of the samples is very high (e.g. $r\approx1$), otherwise we would only retrieve training examples.
\par\medskip
In conclusion, the spectral results derived in this paper have been applied to investigate the retrieval properties of Hopfield-like models, and, even in the worse scenario (where our working assumptions break down), we were able to give a qualitative picture of the processes taking place in associative neural networks while relaxing to fixed points for the neural dynamics.

\section{Acknowledgments}
E.A. and A.F. acknowledge financial support from PNRR MUR project PE0000013-FAIR and from Sapienza University of Rome (RM120172B8066CB0, AR2221815D7192C1, AR1221815EA97525). A.F. has been fully supported by PNRR MUR project PE0000013-FAIR.\\
E.A. and A.F. acknowledge the stimulating research environment provided by the Alan Turing Institute’s Theory and Methods Challenge Fortnights event “Physics-informed Machine Learning”.
\appendix

\section{Proof of Lemma \ref{lem:1}} \label{app:A0}

\begin{proof}[\unskip\nopunct]

	\begin{enumerate}
		\item Differentiating \eqref{eq:what} w.r.t. the dreaming time, we have
		\begin{equation*}
			\begin{split}
				\bb{\dot{J}}&= \frac1{D_N}\bb X^T \frac{d}{dt} \Big(\frac{1+t}{\bb 1+t \bb C}\Big)\bb X= \frac{1}{D_N}\bb X\Big(\frac{1}{\bb 1+t \bb C}-(1+t)\frac{1}{\bb1 +t\bb  C}\bb C\frac{1}{\bb1 +t\bb  C}\Big)\bb X^T=\\&
				=\frac{1}{1+t}\Big(\frac1{D_N} \bb X \frac{1+t}{\bb1+t \bb C}\bb X^T-\frac1{D_N}\bb X\frac{1+t}{\bb1 +t\bb C}\frac{\bb X\bb X^T}{D_N} \frac{1+t}{\bb1+t \bb C}\bb X^T\Big)=\\
				&=\frac{1}{1+t}(\bb J(t)-\bb J(t)^2).
			\end{split}
		\end{equation*}
	
	\item
	Let us start from the eigenvalue problem $\bb J \bb v_\alpha = \lambda_\alpha \bb v_\alpha$, and differentiate w.r.t. the dreaming time:
	\begin{equation*}
		\dot{(\bb J \bb v_\alpha)}= \dot{\bb J}\bb v_\alpha + \bb J \dot{\bb v}_\alpha= \frac{1}{1+t}( \bb J-\bb J^2)\bb v_\alpha + \bb J \dot{\bb v}_\alpha= \frac1{1+t}(\lambda_\alpha-\lambda_\alpha^2)\bb v_\alpha+ \bb J \dot{\bb v}_\alpha.
	\end{equation*}
On the other hand, we have
\begin{equation*}
	\dot{(\bb J\bb v _\alpha)}= \dot \lambda_\alpha \bb v_\alpha+\lambda_\alpha \dot{\bb v}_\alpha.
\end{equation*}
Combining the previous equations, we have
\begin{equation*}
	 \dot \lambda_\alpha \bb v_\alpha+\lambda_\alpha \dot{\bb v}_\alpha=\frac1{1+t}(\lambda_\alpha-\lambda_\alpha^2)\bb v_\alpha+ \bb J \dot{\bb v}_\alpha.
\end{equation*}
Moving the terms proportional to $\bb v_\alpha$ in the l.h.s. and moving those involving $\dot{\bb v}_\alpha$ in the r.h.s., we have
\begin{equation*}
\big(\dot \lambda_\alpha-\frac1{1+t}\lambda_\alpha+\frac1{1+t}\lambda_\alpha ^2\big)\bb v_{\alpha}= (\bb J -\lambda_\alpha )\dot{\bb v}_\alpha.
\end{equation*}
Multiplying on the left by $\bb v_\alpha ^T$, which is a left eigenvector of $\bb J$, the r.h.s. is zero, and -- due to the fact that $\bb v_\alpha\neq0$, we have
\begin{equation}\label{eq:differential_lambda}
	\dot \lambda_\alpha-\frac1{1+t}\lambda_\alpha+\frac1{1+t}\lambda_\alpha ^2= 0.
\end{equation}
The solution of the differential equation is
\begin{equation*}
	\lambda_\alpha (t)= \frac{1+t}{1+t\lambda_\alpha^0}\lambda_\alpha^0,
\end{equation*}
where $\lambda_\alpha^0$ is the generic eigenvalue of the Hebbian coupling matrix $\bb J^0 = \frac1{D_N} \bb X^T \bb X$.

\item Let us first consider the random pattern case. In this setting, the Hebbian coupling matrix $\bb J^0=\frac1N \sum_{\mu=1}^P\xi^\mu_i \xi^\mu_j$ is a positive semidefinite random matrix with rank $P\le N$, so it has $P$ positive eigenvalues. From the theory of Wishart matrices \cite{anderson1958introduction,james1964distributions}, it is known that positive eigenvalues are distinct with probability 1, so $\lambda^0_1 > \dots > \lambda_P^0$, and the eigenvalue $\lambda^0=0$ has degeneracy $N-P$. Since the application $\lambda_\alpha^0\to \lambda_\alpha(t)$ given by Eq. \eqref{eq:lambda_streaming} is injective, the algebraic multiplicity of eigenvalues in the spectrum is preserved for all $t>0$ finite (at $t\to\infty$, all positive eigenvalues concentrate around $\lambda=1$). Let us now consider eigenvectors of $\bb J(t)$ with positive eigenvalue. Starting from
\begin{equation*}
	\dot \lambda_\alpha \bb v_\alpha+\lambda_\alpha \dot{\bb v}_\alpha=\frac1{1+t}(\lambda_\alpha-\lambda_\alpha^2)\bb v_\alpha+ \bb J \dot{\bb v}_\alpha.
\end{equation*}
given in the previous point, and using the differential equation \eqref{eq:differential_lambda}, we get
$$
(\bb J-\lambda_\alpha)\dot {\bb v}_\alpha=0,
$$
so that $\dot{\bb v}_\alpha$ is also eigenvector of $\bb J$ with the same eigenvalue of $\bb v_{\alpha}$. Since for positive eigenvalues, the associated eigenspace is one-dimensional, it follows that $\dot{\bb v}_\alpha = c(t)\bb v_{\alpha},$ whose solution is
\begin{equation}
	\bb v_\alpha(t)= \bb v_\alpha (0) \cdot \exp\int ^t dt' c(t').
\end{equation}
Then, the only effect of dreaming time is the rescaling of the eigenvectors norm. If we take $\bb v_\alpha(t)$ to be the normalized eigenvectors with positive eigenvalues, it follows that $\bb v_\alpha (t)= \bb v_\alpha(0)$, i.e. they do not depend on $t$. For the eigenvalue $\lambda=0$, its eigenspace is $N-P$-dimensional. Denoting $\bb v_0 ^{(1)}(t),\dots,\bb v_0 ^{(N-P)}(t)$ at finite dreaming time $t\ge 0$ the associate eigenvectors, we can write
$$
\bb v_0 ^{(n)}(t)= \sum_{m=1}^{N-P} U_{n,m}(t)\bb v_0 ^{(m)}(0), \quad n=1,\dots,N-P,
$$
which is nothing but a change of basis in the $\lambda=0$ eigenspace. Since we are free to map eigenspaces in themselves without altering the structure of coupling matrix, we can choose $\bb U(t)=\bb 1$ for all $t>0$.
\par\medskip
For the supervised setting the situation is analogous, the only difference being the different structure of the information vectors, whose relevant details are directly encoded in the $t=0$ limit of the coupling matrix $\bb J^0$. For the unsupervised setting, the situation is different. First of all, it is clear that the rank of the matrix will be $\text{min}(N,PM)\equiv N \text{min}(1,\alpha M)$. However, we are mostly interested in the case where the number of examples per class is sufficiently high, i.e. $M \gg 1$ regardless of the value of $\alpha$: in this case, the rank of the coupling matrix $\bb J^0$ would be $N$, then the matrix is full-rank, and all of the eigenvalues would be positive and distinct (for the same reasons of the random pattern case). Thus, positivity and non-degeneracy of the eigenvalues, relation \eqref{eq:lambda_streaming} and stability of eigenvectors w.r.t. dreaming mechanism trivially follows also in this case.
		\end{enumerate}
\end{proof}


\section{Proof of Theorem \ref{thm:1}}\label{app:A}
\begin{proof}[\unskip\nopunct]
	\begin{enumerate}
		\item Let us start again with the basic storing case, where $\bb X = \bb \xi$. As we already said, in this setting the eigenvalue $\lambda=0$ has degeneracy $N-P$, thus the limiting spectral distribution would have a delta around 0 with mass $1-\alpha$. We now focus on positive eigenvalues and start again with the eigenvalue problem $\bb J^0 \bb v_\alpha = \frac1N \bb \xi^T \bb \xi \bb v_\alpha = \lambda_\alpha^0 \bb v_\alpha$. Multiplying on the left by $\bb \xi$, we have
		\begin{equation*}
			\frac1N\bb \xi \bb \xi^T \bb \xi \bb v_\alpha= \bb C  \bb \xi \bb v_\alpha = \lambda_\alpha ^0\bb \xi \bb v_\alpha.
		\end{equation*}
	Thus, positive eigenvalues of the coupling matrix $\bb J^0$ are exactly the eigenvalues of the usual correlation matrix, and the corresponding eigenvector is
	\begin{equation}
		\bb e_\alpha = \frac1{\sqrt{\lambda_\alpha^0 N}}\bb \xi \bb v_\alpha,
	\end{equation}
where the prefactor is needed to ensure the normalization $\lVert \bb e_\alpha \lVert =1$.
	By universality arguments holding for centered patterns with finite variance \cite{Universality-AAP939}, positive eigenvalues of the correlation matrix will be Marchenko-Pastur-distributed with $\text{MP}(\alpha)= \text{MP}(\alpha,1)$, since $\mE{}( {\xi^\mu_i})^2=1$. Since positive eigenvalues have mass $\alpha$, it trivially follows that the empirical spectral distribution $\mu^0_N (\lambda)$ will converge in weak topology to $\mu^0$, in such a way that
	\begin{equation}
		d\mu^0 (\lambda)= (1-\alpha )\delta(\lambda)d\lambda+\alpha d\mu_{\text{MP}}(\lambda),
	\end{equation}
with
$$
d\mu_{\text{MP}}(\lambda)=\frac{1}{2\pi }\frac{\sqrt{(\lambda^0_+-\lambda)(\lambda-\lambda^0_-)}}{\alpha \lambda}d\lambda,
$$
and $\lambda^0_{\pm}= (1\pm\sqrt{\alpha})^2$. Thus, in the random pattern case, $\delta=0$, $\lambda_{\text{peak}}=0$ and $\sigma^2=1$.
\par\medskip
For the supervised case, the situation is similar. Indeed, also in this case the spectral distribution will have a delta peak at $\lambda=0$ with mass $1-\alpha$. For the bulk distribution, the relation between positive eigenvalues of the coupling matrix $\bb J^0$ and the correlation matrix (which is now computed with the empirical means of examples in each class), still holds provided that we replace $\xi^\mu_i \to \bar \xi^\mu_i = \frac1M \sum_A\xi^{\mu,A}_i$. By strong law of large numbers, $\bar \xi^\mu_i \overset{a.s.}\to \mE{\bb\chi}\bar \xi^\mu_i =r \zeta^\mu_i$; this means that $J_{ij}^0\overset{a.s.}\to \frac1{N}\sum_{\mu} \mE{\bb \chi}\bar{\xi}^\mu_i \mE{\bb \chi} \bar \xi^\mu_j= \frac{1}N \sum_\mu (r\zeta^\mu_i )(r \zeta^\mu_j)=r^2 J^{0,\bb\zeta}_{ij}$, where $J_{ij}^{0,\bb \zeta}$ is the Hebbian matrix in the random pattern case built with the ground-truths features, i.e.
\begin{equation*}
	J_{ij}^{0,\bb\zeta}= \frac1N \sum_\mu \zeta^\mu_i \zeta^\mu_j.
\end{equation*}
Now, notice that $\mE{\bb\zeta } ( \zeta^\mu_i)=0$ and $\mE{\bb\zeta}( \zeta^\mu_i)^2=1$ in the $M\to\infty$ limit, eigenvalues of the correlation matrix
$$
\bb C= \frac1{D_N} \bb X \bb X^T,
$$
are distributed according to the Marchenko-Pastur law $\text{MP} (\alpha,r^2)$. Thus, in the supervised case, the empirical spectral distribution of the coupling matrix $\bb J^0$ will converge in weak topology to $\mu^0$, with measure
\begin{equation}
	d\mu ^0 (\lambda)= (1-\alpha)\delta(\lambda)d\lambda+\alpha d\mu_{\text{MP}} (\lambda),
\end{equation}
with
	\begin{equation}
	d\mu_{\text{MP}}(\lambda)=\frac{1}{2\pi r^2}\frac{\sqrt{(\lambda^0_+-\lambda)(\lambda-\lambda^0_-)}}{\alpha \lambda}d\lambda,
\end{equation}
and $\lambda^0_\pm = r^2(1\pm \sqrt \alpha)^2$; thus, in the supervised setting, $\delta=0$, $\lambda_{\text{peak}} =0$ and $\sigma^2= r^2$.\par\medskip
In the unsupervised case, by strong law of large number, for $i\neq j$  we have $ J^0_{ij}\overset{a.s.}\to \mE{\bb \chi}  J_{ij}^0$ as $M\to\infty$, while $J^0_{ii}=\alpha$.  Thus, in this case we can safely replace the coupling matrix $\bb J^0$ with its noise-independent version:
\begin{equation}\label{eq:strong_law_unsup}
	\bb J^{0}\overset{a.s.} \to\alpha(1-r^2)\bb 1+r^2 \bb J^{0,\bb \zeta},
\end{equation}
where $J_{ij}^{0,\bb \zeta}$ is again the Hebbian matrix in the random pattern case built with the ground-truths features. Translating Eq. \eqref{eq:strong_law_unsup} for the eigenvalues of $\bb J^{0}$, we see that the quantity
\begin{equation*}
	\frac{\lambda^0-\alpha(1-r^2)}{r^2}
\end{equation*}
has the same distribution of the eigenvalues of the random pattern case. Then, it follows that, as $M\to\infty$ and in the thermodynamic limit, the empirical spectral distribution $\mu^0_N$ of the unsupervised coupling matrix converges in weak topology $\mu^0$, with
\begin{equation*}
	d\mu^0 (\lambda)=(1-\alpha)\delta(\lambda-\alpha(1-r^2))d\lambda+\alpha d\mu_{\text{MP}} (\lambda),
\end{equation*}
with
\begin{equation*}
	d\mu_{\text{MP}} (\lambda)=\frac{1}{2\pi r^2}\frac{\sqrt{(\lambda^0_+-\lambda)(\lambda-\lambda_-^0)}}{\alpha (\lambda-\alpha (1-r^2))}d\lambda,
\end{equation*}
with $\lambda^0_\pm = r^2 (1\pm \sqrt \alpha)^2+\alpha(1-r^2)$. Thus, in the unsupervised case, we have $\delta=\lambda_{\text{peak}}=\alpha(1-r^2)$ and $\sigma^2=r^2$.
\item The proof works by reverting Eq. \eqref{eq:lambda_streaming}, expressing $\lambda_\alpha ^0$ as a function of $\lambda_\alpha(t)$. Thus, in the thermodynamic limit (and eventually for $M\to\infty$), the empirical spectral distribution $\mu_N ^t$ will converge in weak topology to $\mu^t$, with the latter determined by the fact that the quantity
$$
\frac{\lambda}{1+t(1-\lambda)},
$$
will be equal in distribution to $\lambda^0$, regardless of the setting under consideration.
	\end{enumerate}
\end{proof}


\section{Proof of Proposition \ref{prop:1}}\label{app:B}

\begin{proof}[\unskip\nopunct]
	First of all, let us notice that, since in the $M\to\infty$ limit, $\bb J^{0} \overset{a.s.}\to r^2 \bb J^{0,\bb\zeta}$ in the supervised setting and $\bb J^{0} \overset{a.s.}\to\alpha (1-r^2)\bb 1+ r^2 \bb J^{0,\bb\zeta}$ in the unsupervised one, and the fact that eigenvectors can be chosen so that they do not depend on $t$, $\bb J$ and $\bb J^{\bb\zeta}$  (the latter being the dreaming coupling matrix in the random pattern case built with the ground-truths) have common eigenvectors a.s., so they can be simultaneously diagonalized with the transformation $\bb J \to \bb U \bb D \bb U^{-1}$ with the same matrix $\bb U$. Because of these simple relations between the (un)supervised coupling matrix and the corresponding ground-truth version, a functional relation between eigenvalues can be derived. For example, the generic eigenvalue $\lambda^s(t)$ of the supervised coupling matrix is related to the corresponding eigenvalue of $\bb J^0$ through Eq. \eqref{eq:lambda_streaming}. In the $M\to\infty$ limit, $\lambda^s_0\to r^2 \lambda_0$, where $\lambda _0$ is the corresponding eigenvalue of $\bb J^{0,\bb\zeta}$, thus
	$$
	\lambda^s(t)= \frac{(1+t)\lambda^s_0}{1+t\lambda^s_0}= \frac{(1+t)r^2 \lambda_0}{1+tr^2 \lambda_0}.
	$$
	Finally, one can re-express $\lambda_0$ in terms of $\lambda(t)$ being the eigenvalue of $\bb J^{\bb\zeta}(t)$ the coupling matrix of the ground-truths by reverting Eq. \eqref{eq:lambda_streaming}, giving us
	$$
	\lambda^s(t)=\frac{\lambda (t) r^2 (t+1)}{\lambda(t)  \left(r^2-1\right) t+t+1}=	f^s_{r,t}(\lambda(t)).
	$$
	Clearly, with the same procedure, one finds that the functional relation for the eigenvalues of the unsupervised coupling matrix is $\lambda^u (t)= f^u_{r,t}(\lambda(t))$. With these results, we find for the SE the expression
	\begin{equation}
		\delta ^{s,u}(\alpha,r,t)= \frac 1N \text{Tr} (\bb J^{\bb\zeta}- \bb J ^{s,u}(t))^2= \frac1N \sum_\alpha (\lambda_\alpha (t)-f^{s,u}_{r,t}(\lambda_\alpha (t)))^2\to  \int (\lambda - f^{s,u}_{r,t}(\lambda))^2d\mu^t(\lambda),
	\end{equation}
where $\to$ stands for convergence in probability in the thermodynamic limit.
\end{proof}

\begin{Rmk}
	Because of the structure of the limiting spectral distribution of the dreaming coupling matrix $d\mu^t$, the SE can be rewritten as
	$$
		\delta^{s}(\alpha,r,t)= \alpha  \int (\lambda - f^{s}_{r,t}(\lambda))^2d\mu^t_{\text{bulk}}(\lambda),
	$$
	for the supervised setting, and
	$$
	\delta^{u}(\alpha,r,t)=(1-\alpha)\Big[\frac{\alpha  \left(r^2-1\right) (t+1)}{\alpha  \left(r^2-1\right) t-1}\Big]^2+\alpha  \int (\lambda - f^{u}_{r,t}(\lambda))^2d\mu^t_{\text{bulk}}(\lambda).
	$$
	In the last expression, the constant contribution comes from the presence of the delta peak located at non-vanishing eigenvalue for the coupling matrix in the unsupervised setting.
\end{Rmk}

\section{Proof of Proposition \ref{prop:pattern_computations} and details on the GA} \label{app:D}

\begin{proof}[\unskip\nopunct]
	The proof works by explicit computation of the empirical moments. Let us start with the pattern stability. Under the GA assumption in the thermodynamic limit, and since patterns are equivalent (so that we can take the average also average the index $\mu$), we can estimate
	\begin{eqnarray}
		\mu_1 &=& \frac1{NP}\sum_{i\mu}\Delta_i(\bb\xi^\mu) ,\\
		\mu_2 &=& \frac1{NP}\sum_{i\mu} \Delta_i(\bb\xi^\mu)^2.
	\end{eqnarray}
For the first quantity, we have
\begin{equation}
	\begin{split}
		\mu_1 &= \frac1{NP}\sum_{i\mu} J_{ij}(t)\xi^\mu _i \xi^\mu_j = \frac1{\alpha N}\sum_{ij}J_{ij}(t)J_{ij}(0)= \frac1{\alpha N}\text{Tr} \bb J (t)\bb J(0).
	\end{split}
\end{equation}
Now, $\bb J(t)$ and $\bb J^0$ are simultaneously diagonalizable, thus
\begin{equation}
	\mu_1 = \frac1{\alpha N}\sum_\alpha \lambda_\alpha (t)\lambda_\alpha ^0 \underset{TDL}\to \frac1\alpha \int \frac{\lambda^2}{1+t(1-\lambda)}d\mu^t(\lambda),
\end{equation}
where we expressed $\lambda_\alpha^0$ as a function of $\lambda_\alpha (t)$ by reverting Eq. \eqref{eq:lambda_streaming}. Analogously, for the second moment
\begin{equation}
	\begin{split}
		\mu_2 &= \frac1{NP}\sum_{i\mu jk}J_{ij}(t)J_{ik}(t)\xi^\mu_j\xi^\mu_k= \frac1{\alpha N}\sum_{ijk} J_{ij}(t)J_{ik}(t)J_{jk}(0)=\\
		&=\frac 1{\alpha N}\text{Tr} \bb J(t)^2\bb J(0)\underset{TDL}\to \frac1\alpha \int\frac{\lambda^3}{1+t(1-\lambda)}d\mu^t(\lambda).
	\end{split}
\end{equation}
As for the attractiveness, the computations follows the same lines, provided that we use \eqref{eq:pattern_attract_mod} as definition, and under the GA we average the moments w.r.t. $\bb\eta$, and noticing that $\mE{\bb \eta} \eta_i =p$ and $\mE{\bb\eta}\eta_j \eta_k = (1-p^2)\delta_{jk} +p^2$.
\end{proof}

\begin{Rmk}
	Notice that we can recast everything in terms of integrals of usual Marchenko-Pastur distribution with scale parameter $\alpha<1$. Indeed, by using spectral decomposition of the coupling matrix we can write the first empirical moment of the stability as
		\begin{equation}\label{eq:tgg1_c1}
		\begin{split}
			\mu_1& = \frac1{NP}\mE{\bb\xi}\sum_{ij\mu} J_{ij}\xi^\mu_i \xi^\mu_j = \frac1{NP}\mE{\bb\xi}\sum_{\alpha i j\mu} \lambda_\alpha v^i_\alpha v^j_\alpha \xi^\mu_i \xi^\mu_j= \frac1{NP}\mE{\bb\xi}\sum_{\alpha\mu}\lambda_\alpha \Big(\sum_i v^i_\alpha \xi^\mu_i\Big)\Big(\sum_j v^j_\alpha \xi^\mu_j\Big)=\\
			&=\frac 1{NP}\sum_{\alpha\mu} \lambda_\alpha (\sqrt{\lambda_\alpha ^0 N } e^\mu_\alpha)^2= \frac1P \sum_{\alpha} \lambda _\alpha \lambda_\alpha^0 \underset{TDL}\to\int \frac{(1+t)\lambda^2}{1+t\lambda}d\mu_{\text{MP}}(\lambda),
		\end{split}
	\end{equation}
	where we used $\bb e _\alpha = (\lambda_\alpha^0 N)^{-1/2}\bb \xi \bb v_\alpha$ such that $\sum_\mu (e^\mu_\alpha)^2=1$ are the eigenvectors of the correlation matrix, and the fact that $\lambda_\alpha = (1+t)\lambda^0_\alpha/(1+t\lambda^0_\alpha)$ and that the coupling matrices have only $P$ positive eigenvalues. Similarly, for the second moment
	\begin{equation}\label{eq:tgg1_c2}
		\mu_2 \underset{TDL} \to \int \frac{(1+t)^2\lambda^3}{(1+t\lambda)^2}d\mu_{\text{MP}}(\lambda).
	\end{equation}
\end{Rmk}
\begin{Rmk}
	In order to check the validity of the GA, we consider the third centered moment of the attractiveness, which in the thermodynamic limit can be approximated as
	\begin{equation}
		\mE{\bb\eta}(\Delta_i ^\mu -\mE{\bb\eta} \Delta_i ^\mu)^3\underset{TDL}\sim \frac1{NP}\mE{\bb\xi}\sum_{i\mu jkl}J_{ij}J_{ik}J_{il}\xi^\mu_i \xi^\mu_j \xi^\mu_k \xi^\mu_l \mE{\bb\eta} (\eta_j-p) (\eta_k-p)(\eta_l-p).
	\end{equation}
	Noticing that $\mE{\bb\eta} (\eta_j-p) (\eta_k-p)(\eta_l-p)= 2p(1-p^2)\delta_{jk}\delta_{kl}$, it follows that
	\begin{equation}
		\mE{\bb\eta}(\Delta_i ^\mu -\mE{\bb\eta} \Delta_i ^\mu)^3=\frac{2p(1-p^2)}{NP}\mE{\bb\xi}\sum_{i\mu j} J_{ij}^3 \xi^\mu_i \xi^\mu_j .
	\end{equation}
	We can thus bound the third centered moment as
	\begin{equation*}
		\begin{split}
		 \vert \mE{\bb\eta}(\Delta_i ^\mu -\mE{\bb\eta} \Delta_i ^\mu)^3\vert& \le \frac{2p(1-p^2)}{NP}\mE{\bb\xi}\sum_{i\mu j} \vert J_{ij}\vert ^3=\frac{2p(1-p^2)}{N}\mE{\bb\xi}\sum_{i j} \vert J_{ij}\vert ^3\le \frac{2p(1-p^2)}{N}\mE{\bb\xi}\sum_{i j} \vert J_{ij}\vert ^2=\\&=\frac{2p(1-p^2)}{N}\mE{\bb\xi}\text{Tr}\bb J^2\to 2p(1-p^2)\int \lambda^2 d\mu^t(\lambda),
		\end{split}
	\end{equation*}
	where we used the fact that $\vert J_{ij}\vert \le1$. Then, a necessary condition for the third centered moment to be close to zero is
	$$
	2\alpha p(1-p^2)\int \lambda^2 d\mu^t(\lambda)\ll1.
	$$
	
\end{Rmk}

\section{Proof of proposition \ref{prop:un_sup}} \label{app:E}
\begin{proof}[\unskip\nopunct]
    \begin{enumerate}
        \item In the supervised setting, the empirical first moment of the attractiveness is 
        \begin{equation*}
            \begin{split}
                \mu_1 &=\frac1{NP}\mE{\bb\chi}\sum_{i\mu j}J_{ij}(t)\chi _j \zeta^\mu_i \zeta^\mu_j =\frac{r}{NP}\sum_{i\mu j}J_{ij}(t)\zeta^\mu_i \zeta^\mu_j= \frac{r}{\alpha N}\sum_{ij} J_{ij}(t) J^{0,\bb\zeta}_{ij}=\frac r{\alpha N} \text{Tr}\, \bb J(t) \bb J^{0,\bb\zeta},
            \end{split}
        \end{equation*}
        where $\bb J^{0,\bb\zeta}$ is the Hebbian matrix in the archetype-setting with ground-truths $\bb\zeta$. In the $M\to\infty$ limit, since $\bb J^0\overset{a.s.}\to r^2 \bb J^{0,\bb\zeta}$, we can safely write
        \begin{equation*}
            \begin{split}
                \mu_1 &= \frac{r}{\alpha N}\sum_{ij} J_{ij}(t) \frac1{r^2} J^{0}_{ij}=  \frac{1}{\alpha r N}\text{Tr}\, \bb J(t) \bb J^0\to \frac1{\alpha r}\int \frac{\lambda^2 }{1+t(1-\lambda)}d\mu^t _s (\lambda),
            \end{split}
        \end{equation*}
        in the thermodynamic limit. For the empirical second moment, we have
        \begin{equation*}
            \begin{split}
                \mu_2 &=\frac1{NP}\mE{\bb\chi}\sum_{i\mu jk}J_{ij}(t)J_{ik}(t)\chi _j\chi _k \zeta^\mu_j \zeta^\mu_k =\frac{1-r^2}{N}\sum_{i j}J_{ij}(t)^2+\frac{r^2}{NP} \sum_{i\mu jk}J_{ij}(t)J_{ik}(t)  \zeta^\mu_j \zeta^\mu_k =\\
                &=\frac{1-r^2}{N}\text{Tr}\, \bb J(t)^2 +\frac{r^2}{\alpha N}\text{Tr}\, \bb J(t)^2 \bb J^{0,\bb\zeta}=\frac{1-r^2}{N}\text{Tr}\, \bb J(t)^2 +\frac{1}{\alpha N}\text{Tr}\, \bb J(t)^2 \bb J^{0}\to\\
                &\to (1-r^2)\int \lambda^2 d\mu^t _s (\lambda)+ \frac{1}{\alpha} \int \frac{\lambda^3 }{1+t(1-\lambda)}d\mu^t _s(\lambda),
            \end{split}
        \end{equation*}
        in the thermodynamic limit.

        \item In the unsupervised setting, we still have
        \begin{equation*}
            \mu_1 = \frac{r}{\alpha N}\text{Tr}\,\bb J (t)\bb  J^{0,\bb\zeta},
        \end{equation*}
        as in the supervised case. However, in the setting under consideration, in the $M\to\infty$ limit we have
        $$
        \bb J^{0,\bb\zeta}\equiv \frac{1}{r^2}(\bb J^0 -\alpha (1-r^2) \bb 1).
        $$
        Thus, we have
        \begin{equation*}
            \begin{split}
                \mu_1 &= \frac 1{\alpha r N} \text{Tr}\, \bb J(t) \bb J^0-\frac{1-r^2}{r N}\text{Tr}\, \bb J(t)\to\frac1{\alpha r }\int \frac{\lambda^2}{1+t(1-\lambda)}d\mu^t_u(\lambda)-\frac{1-r^2}{r}\int \lambda d\mu^t_u (\lambda),
            \end{split}
        \end{equation*}
        in the thermodynamic limit. Analogously, for the empirical second moment we have
        \begin{equation*}
            \begin{split}
                \mu_2 &=\frac{1-r^2}{N}\text{Tr}\, \bb J(t)^2 +\frac{r^2}{\alpha N}\text{Tr}\, \bb J(t)^2 \bb J^{0,\bb\zeta}=\frac{1-r^2}N \text{Tr}\bb J(t)^2 +\frac1{\alpha N } \text{Tr}\, \bb J(t)^2 \bb J^0 -\frac{1-r^2}{N}\text{Tr}\, \bb J(t)^2=\\
                &=\frac1{\alpha N } \text{Tr}\, \bb J(t)^2 \bb J^0\to \frac{1}{\alpha}\int \frac{\lambda^3}{1+t(1-\lambda)}d\mu^t_u (\lambda),
            \end{split}
        \end{equation*}
        in the thermodynamic limit.
    \end{enumerate}
\end{proof}

\bibliographystyle{ieeetr}

\begin{thebibliography}{10}

\bibitem{hopfield1982neural}
J.~J. Hopfield, ``Neural networks and physical systems with emergent collective
  computational abilities.,'' {\em Proceedings of the national academy of
  sciences}, vol.~79, no.~8, pp.~2554--2558, 1982.

\bibitem{hopfield1985neural}
J.~J. Hopfield and D.~W. Tank, ``“neural” computation of decisions in
  optimization problems,'' {\em Biological cybernetics}, vol.~52, no.~3,
  pp.~141--152, 1985.

\bibitem{Hebb-1949}
D.~Hebb, {\em The Organization of Behavior: A Neuropsychological Theory.}
\newblock New York, NY: John Wiley \& Sons, 1949.

\bibitem{Amit}
D.~Amit, {\em Modeling brain function: The world of attractor neural networks}.
\newblock Cambridge university press, 1989.

\bibitem{AGS1}
D.~J. Amit, H.~Gutfreund, and H.~Sompolinsky, ``Storing infinite numbers of
  patterns in a spin-glass model of neural networks,'' {\em Phys. Rev. Lett.},
  vol.~55, pp.~1530--1533, Sep 1985.

\bibitem{AGS2}
D.~J. Amit, H.~Gutfreund, and H.~Sompolinsky, ``Statistical mechanics of neural
  networks near saturation,'' {\em Annals of physics}, vol.~173, no.~1,
  pp.~30--67, 1987.

\bibitem{bovier1995large}
A.~Bovier, V.~Gayrard, and P.~Picco, ``Large deviation principles for the
  hopfield model and the kac-hopfield model,'' {\em Probability theory and
  related fields}, vol.~101, no.~4, pp.~511--546, 1995.

\bibitem{bovier1996almost}
A.~Bovier and V.~Gayrard, ``An almost sure large deviation principle for the
  hopfield model,'' {\em The annals of probability}, vol.~24, no.~3,
  pp.~1444--1475, 1996.

\bibitem{barra2012glassy}
A.~Barra, G.~Genovese, F.~Guerra, and D.~Tantari, ``How glassy are neural
  networks?,'' {\em Journal of Statistical Mechanics: Theory and Experiment},
  vol.~2012, no.~07, p.~P07009, 2012.

\bibitem{agliari2020generalized}
E.~Agliari, F.~Alemanno, A.~Barra, and A.~Fachechi, ``Generalized guerra’s
  interpolation schemes for dense associative neural networks,'' {\em Neural
  Networks}, vol.~128, pp.~254--267, 2020.

\bibitem{AABO-JPA2020}
E.~Agliari, L.~Albanese, A.~Barra, and G.~Ottaviani, ``Replica symmetry
  breaking in neural networks: A few steps toward rigorous results,'' {\em
  Journal of Physics A: Mathematical and Theoretical}, vol.~53, 2020.

\bibitem{bovier1999sharp}
A.~Bovier, ``Sharp upper bounds on perfect retrieval in the hopfield model,''
  {\em Journal of applied probability}, vol.~36, no.~3, pp.~941--950, 1999.

\bibitem{feng2001critical}
J.~Feng, M.~Shcherbina, and B.~Tirozzi, ``On the critical capacity of the
  hopfield model,'' {\em Communications in Mathematical Physics}, vol.~216,
  pp.~139--177, 2001.

\bibitem{loukianova1997lower}
D.~Loukianova, ``Lower bounds on the restitution error in the hopfield model,''
  {\em Probability theory and related fields}, vol.~107, no.~2, pp.~161--176,
  1997.

\bibitem{newman1988memory}
C.~M. Newman, ``Memory capacity in neural network models: Rigorous lower
  bounds,'' {\em Neural Networks}, vol.~1, no.~3, pp.~223--238, 1988.

\bibitem{lowe1998storage}
M.~L{\"o}we, ``On the storage capacity of hopfield models with correlated
  patterns,'' {\em The Annals of Applied Probability}, vol.~8, no.~4,
  pp.~1216--1250, 1998.

\bibitem{bovier1992rigorous}
A.~Bovier and V.~Gayrard, ``Rigorous bounds on the storage capacity of the
  dilute hopfield model,'' {\em Journal of statistical physics}, vol.~69,
  pp.~597--627, 1992.

\bibitem{bovier1993rigorous}
A.~Bovier and V.~Gayrard, ``Rigorous results on the thermodynamics of the
  dilute hopfield model,'' {\em Journal of statistical physics}, vol.~72,
  pp.~79--112, 1993.

\bibitem{baldi1987number}
P.~Baldi and S.~S. Venkatesh, ``Number of stable points for spin-glasses and
  neural networks of higher orders,'' {\em Physical Review Letters}, vol.~58,
  no.~9, p.~913, 1987.

\bibitem{bovier2001spin}
A.~Bovier and B.~Niederhauser, ``The spin-glass phase-transition in the
  hopfield model with $ p $-spin interactions,'' {\em Advances in Theoretical
  and Mathematical Physics}, vol.~5, no.~6, pp.~1001--1046, 2001.

\bibitem{Gardner-JPA1988}
E.~Gardner, ``The space of interactions in neural network models,'' {\em
  Journal of Physics A}, vol.~21, no.~1, p.~257, 1988.

\bibitem{hopfield1983unlearning}
J.~J. Hopfield, D.~I. Feinstein, and R.~G. Palmer, ``‘unlearning’has a
  stabilizing effect in collective memories,'' {\em Nature}, vol.~304,
  no.~5922, pp.~158--159, 1983.

\bibitem{personnaz1985information}
L.~Personnaz, I.~Guyon, and G.~Dreyfus, ``Information storage and retrieval in
  spin-glass like neural networks,'' {\em Journal de Physique Lettres},
  vol.~46, no.~8, pp.~359--365, 1985.

\bibitem{kanter1987associative}
I.~Kanter and H.~Sompolinsky, ``Associative recall of memory without errors,''
  {\em Physical Review A}, vol.~35, no.~1, p.~380, 1987.

\bibitem{plakhov1992modified}
A.~Plakhov and S.~Semenov, ``The modified unlearning procedure for enhancing
  storage capacity in hopfield network,'' in {\em [Proceedings] 1992 RNNS/IEEE
  Symposium on Neuroinformatics and Neurocomputers}, pp.~242--251, IEEE, 1992.

\bibitem{plakhov1994converging}
A.~Y. Plakhov, ``The converging unlearning algorithm for the hopfield neural
  network: optimal strategy,'' in {\em Proceedings of the 12th IAPR
  International Conference on Pattern Recognition, Vol. 3-Conference C: Signal
  Processing (Cat. No. 94CH3440-5)}, vol.~2, pp.~104--106, IEEE, 1994.

\bibitem{plakhov1995convergent}
A.~Y. Plakhov, S.~A. Semenov, and I.~B. Shuvalova, ``Convergent unlearning
  algorithm for the hopfield neural network,'' in {\em Proceedings 1995 Second
  New Zealand International Two-Stream Conference on Artificial Neural Networks
  and Expert Systems}, pp.~30--33, IEEE, 1995.

\bibitem{van1997hebbian}
J.~Van~Hemmen, ``Hebbian learning, its correlation catastrophe, and
  unlearning,'' {\em Network: Computation in Neural Systems}, vol.~8, no.~3,
  p.~V1, 1997.

\bibitem{horas1998unlearning}
J.~A. Horas and P.~M. Pasinetti, ``On the unlearning procedure yielding a
  high-performance associative memory neural network,'' {\em Journal of Physics
  A: Mathematical and General}, vol.~31, no.~25, p.~L463, 1998.

\bibitem{dotsenko1991statistical}
V.~Dotsenko, N.~Yarunin, and E.~Dorotheyev, ``Statistical mechanics of
  hopfield-like neural networks with modified interactions,'' {\em Journal of
  Physics A: Mathematical and General}, vol.~24, no.~10, p.~2419, 1991.

\bibitem{dotsenko1991replica}
V.~Dotsenko and B.~Tirozzi, ``Replica symmetry breaking in neural networks with
  modified pseudo-inverse interactions,'' {\em Journal of Physics A:
  Mathematical and General}, vol.~24, no.~21, p.~5163, 1991.

\bibitem{fachechi2019dreaming}
A.~Fachechi, E.~Agliari, and A.~Barra, ``Dreaming neural networks: forgetting
  spurious memories and reinforcing pure ones,'' {\em Neural Networks},
  vol.~112, pp.~24--40, 2019.

\bibitem{agliari2019dreaming}
E.~Agliari, F.~Alemanno, A.~Barra, and A.~Fachechi, ``Dreaming neural networks:
  rigorous results,'' {\em Journal of Statistical Mechanics: Theory and
  Experiment}, vol.~2019, no.~8, p.~083503, 2019.

\bibitem{fachechi2022outperforming}
A.~Fachechi, A.~Barra, E.~Agliari, and F.~Alemanno, ``Outperforming rbm
  feature-extraction capabilities by “dreaming” mechanism,'' {\em IEEE
  Transactions on Neural Networks and Learning Systems}, 2022.

\bibitem{Fontanari-1990}
J.~Fontanari, ``Generalization in a {H}opfield network,'' {\em Journal of
  Physics France}, vol.~51, pp.~2421--2430, 1990.

\bibitem{AABD-NN2022}
E.~Agliari, F.~Alemanno, A.~Barra, and G.~De~Marzo, ``The emergence of a
  concept in shallow neural networks,'' {\em Neural Networks}, vol.~148,
  pp.~232--253, 2022.

\bibitem{AAKBA-EPL2023}
M.~Aquaro, F.~Alemanno, I.~Kanter, A.~Barra, and E.~Agliari, ``Supervised
  {H}ebbian learning,'' {\em Europhysics Letters - Perspective}, vol.~141,
  p.~11001, 2023.

\bibitem{benedetti2023eigenvector}
M.~Benedetti, L.~Carillo, E.~Marinari, and M.~M{\'e}zard, ``Eigenvector
  dreaming,'' {\em arXiv preprint arXiv:2308.13445}, 2023.

\bibitem{regularizationdreaming}
E.~Agliari, M.~Aquaro, F.~Alemanno, and A.~Fachechi, ``Regularization,
  early-stopping and dreaming: a hopfield-like setup to address generalization
  and overfitting,'' {\em arXiv preprint}, p.~2308.01421, 2023.

\bibitem{leonelli2021effective}
F.~E. Leonelli, E.~Agliari, L.~Albanese, and A.~Barra, ``On the effective
  initialisation for restricted {B}oltzmann machines via duality with
  {H}opfield model,'' {\em Neural Networks}, vol.~143, pp.~314--326, 2021.

\bibitem{kosterlitz1976spherical}
J.~M. Kosterlitz, D.~J. Thouless, and R.~C. Jones, ``Spherical model of a
  spin-glass,'' {\em Physical Review Letters}, vol.~36, no.~20, p.~1217, 1976.

\bibitem{galluccio1998rational}
S.~Galluccio, J.-P. Bouchaud, and M.~Potters, ``Rational decisions, random
  matrices and spin glasses,'' {\em Physica A: Statistical Mechanics and its
  Applications}, vol.~259, no.~3-4, pp.~449--456, 1998.

\bibitem{auffinger2013random}
A.~Auffinger, G.~B. Arous, and J.~{\v{C}}ern{\`y}, ``Random matrices and
  complexity of spin glasses,'' {\em Communications on Pure and Applied
  Mathematics}, vol.~66, no.~2, pp.~165--201, 2013.

\bibitem{zhu2016inverse}
L.~Zhu and W.-w. Xu, ``The inverse eigenvalue problem of structured matrices
  from the design of hopfield neural networks,'' {\em Applied Mathematics and
  Computation}, vol.~273, pp.~1--7, 2016.

\bibitem{pennington2017geometry}
J.~Pennington and Y.~Bahri, ``Geometry of neural network loss surfaces via
  random matrix theory,'' in {\em International conference on machine
  learning}, pp.~2798--2806, PMLR, 2017.

\bibitem{mai2018random}
X.~Mai and R.~Couillet, ``A random matrix analysis and improvement of
  semi-supervised learning for large dimensional data,'' {\em The Journal of
  Machine Learning Research}, vol.~19, no.~1, pp.~3074--3100, 2018.

\bibitem{liao2018dynamics}
Z.~Liao and R.~Couillet, ``The dynamics of learning: A random matrix
  approach,'' in {\em International Conference on Machine Learning},
  pp.~3072--3081, PMLR, 2018.

\bibitem{seddik2020random}
M.~E.~A. Seddik, C.~Louart, M.~Tamaazousti, and R.~Couillet, ``Random matrix
  theory proves that deep learning representations of gan-data behave as
  gaussian mixtures,'' in {\em International Conference on Machine Learning},
  pp.~8573--8582, PMLR, 2020.

\bibitem{zhou2021eigenvalue}
J.~Zhou, Z.~Jiang, T.~Hou, Z.~Chen, K.~M. Wong, and H.~Huang, ``Eigenvalue
  spectrum of neural networks with arbitrary hebbian length,'' {\em Physical
  Review E}, vol.~104, no.~6, p.~064307, 2021.

\bibitem{couillet2022random}
R.~Couillet and Z.~Liao, {\em Random matrix methods for machine learning}.
\newblock Cambridge University Press, 2022.

\bibitem{granziol2022random}
D.~Granziol and N.~Baskerville, ``A random matrix theory approach to damping in
  deep learning,'' {\em Journal of Physics: Complexity}, vol.~3, no.~2,
  p.~024001, 2022.

\bibitem{barbier2023fundamental}
J.~Barbier, F.~Camilli, M.~Mondelli, and M.~S{\'a}enz, ``Fundamental limits in
  structured principal component analysis and how to reach them,'' {\em
  Proceedings of the National Academy of Sciences}, vol.~120, no.~30,
  p.~e2302028120, 2023.

\bibitem{LenkaFlorent}
L.~Zdeborová and F.~Krzakala, ``Statistical physics of inference: thresholds
  and algorithms,'' {\em Advances in Physics}, vol.~65, no.~5, pp.~453--552,
  2016.

\bibitem{LenkaJPA}
E.~Agliari, A.~Barra, P.~Sollich, and L.~Zdeborova, ``Machine learning and
  statistical physics: theory, inspiration, application,'' {\em J. Phys. A:
  Math. and Theor.}, vol.~Special, 2020.

\bibitem{zanin23}
P.~Zanin and N.~Caticha, ``Interacting dreaming neural networks,'' {\em Journal
  of Statistical Mechanics}, vol.~2034, p.~043401, 2023.

\bibitem{serricchio23}
L.~Serricchio, C.~Chilin, D.~Bocchi, R.~Marino, M.~Negri, C.~Cammarota, and
  F.~Ricci-Tersenghi, ``Daydreaming hopfield networks and their surprising
  effectiveness on correlated data,'' in {\em Associative Memory {\&} Hopfield
  Networks in 2023}, 2023.

\bibitem{camilli23}
F.~Camilli and M.~M\'ezard, ``The decimation scheme for symmetric matrix
  factorization,'' {\em arxiv:2307.16564}, 2023.

\bibitem{VCMZ-2023}
E. Ventura, S. Cocco, R. Monasson, and F. Zamponi, ``Unlearning regularization for Boltzmann Machines,'' {\em arxiv:2311.09418}, 2023.

\bibitem{Kohonen-1972}
T.~Kohonen and M.~Ruohonen, ``Representation of {A}ssociated {D}ata by {M}atrix
  {O}perators,'' {\em IEEE Transaztions on Computers}, 1973.

\bibitem{AlbertIEEE}
A.~Fachechi, A.~Barra, E.~Agliari, and F.~Alemanno, ``Outperforming {RBM}
  feature-extraction capabilities by ``dreaming'' mechanism,'' {\em IEEE
  Transactions on Neural Networks and Learning Systems}, pp.~1--10, 6 2022.

\bibitem{pallara}
L.~Albanese, A.~Barra, P.~Bianco, F.~Durante, and D.~Pallara, ``Hebbian
  learning from first principles,'' {\em arXiv:2401.07110}, 2024.

\bibitem{lad}
E.~Agliari, F.~Alemanno, M.~Aquaro, A.~Barra, F.~Durante, and I.~Kanter,
  ``Hebbian dreaming for small datasets,'' {\em arXiv:2204.07954}.

\bibitem{MP}
V.~A. {Mar{\v{c}}enko} and L.~A. {Pastur}, ``{Distribution of Eigenvalues for
  Some Sets of Random Matrices},'' {\em Sbornik: Mathematics}, vol.~1,
  pp.~457--483, Apr. 1967.

\bibitem{Kanter-Sompolinsky-1987}
I.~Kanter and H.~Sompolinsky, ``Associative recall of memory without errors,''
  {\em Physical Review A}, vol.~35, no.~1, p.~380, 1987.

\bibitem{sommers-PRL1988}
H.~J. Sommers, A.~Crisanti, H.~Sompolinsky, and Y.~Stein, ``Spectrum of large
  random asymmetric matrices,'' {\em Phys. Rev. Lett.}, vol.~60,
  pp.~1895--1898, May 1988.

\bibitem{galluccio-physa1998}
S.~Galluccio, J.-P. Bouchaud, and M.~Potters, ``Rational decisions, random
  matrices and spin glasses,'' {\em Physica A: Statistical Mechanics and its
  Applications}, vol.~259, no.~3, pp.~449--456, 1998.

\bibitem{kanaka-PRL2006}
K.~Rajan and L.~F. Abbott, ``Eigenvalue spectra of random matrices for neural
  networks,'' {\em Phys. Rev. Lett.}, vol.~97, p.~188104, Nov 2006.

\bibitem{castillo-PRE2008}
T.~Rogers, I.~P. Castillo, R.~K\"uhn, and K.~Takeda, ``Cavity approach to the
  spectral density of sparse symmetric random matrices,'' {\em Phys. Rev. E},
  vol.~78, p.~031116, Sep 2008.

\bibitem{AABF-JPA2019}
E.~Agliari, F.~Alemanno, A.~Barra, and A.~Fachechi, ``On the marchenko–pastur
  law in analog bipartite spin-glasses*,'' {\em Journal of Physics A:
  Mathematical and Theoretical}, vol.~52, p.~254002, may 2019.

\bibitem{CD}
I.~Sutskever and T.~Tieleman, ``On the convergence properties of contrastive
  divergence,'' {\em Journal of Machine Learning Research}, vol.~9, p.~9, 2010.

\bibitem{Rocchi_2017}
J.~Rocchi, D.~Saad, and D.~Tantari, ``High storage capacity in the hopfield
  model with auto-interactions—stability analysis,'' {\em Journal of Physics
  A: Mathematical and Theoretical}, vol.~50, p.~465001, oct 2017.

\bibitem{mceliece1987capacity}
R.~McEliece, E.~Posner, E.~Rodemich, and S.~Venkatesh, ``The capacity of the
  hopfield associative memory,'' {\em IEEE transactions on Information Theory},
  vol.~33, no.~4, pp.~461--482, 1987.

\bibitem{anderson1958introduction}
T.~W. Anderson, {\em An introduction to multivariate statistical analysis},
  vol.~2.
\newblock Wiley New York, 1958.

\bibitem{james1964distributions}
A.~T. James, ``Distributions of matrix variates and latent roots derived from
  normal samples,'' {\em The Annals of Mathematical Statistics}, vol.~35,
  no.~2, pp.~475--501, 1964.

\bibitem{Universality-AAP939}
N.~S. Pillai and J.~Yin, ``{Universality of covariance matrices},'' {\em The
  Annals of Applied Probability}, vol.~24, no.~3, pp.~935 -- 1001, 2014.

\end{thebibliography}

\end{document}